\documentclass{aa}
\usepackage{graphicx}
\usepackage{txfonts}
\usepackage{natbib}
\bibpunct{(}{)}{;}{a}{}{,} 
\usepackage{color}
\usepackage[colorlinks=true,linkcolor=black,citecolor=blue,urlcolor=blue]{hyperref}

\begin{document} 
   \title{The chemistry of episodic accretion in embedded objects.}
   \subtitle{2D radiation thermo-chemical models of the post-burst phase.}

   \author{Ch. Rab\inst{1}   
   \and V. Elbakyan\inst{2}
   \and E. Vorobyov\inst{3,2,1} 
   \and M. Güdel\inst{1} 
   \and O. Dionatos\inst{1}
   \and M. Audard\inst{4}
   \and I. Kamp\inst{5}
   \and W.-F. Thi\inst{6}
   \and P. Woitke\inst{7}
   \and A. Postel\inst{4}
          }

  \titlerunning{}
  \authorrunning{Ch. Rab et al.}

\institute{University of Vienna, Dept. of Astrophysics, T\"urkenschanzstr. 17, 1180 Wien, Austria \\
         \email{christian.rab@univie.ac.at}  
\and  Research Institute of Physics, Southern Federal University, Stachki 194,
Rostov-on-Don, 344090, Russia
\and Institute of Fluid Mechanics and Heat Transfer, TU Wien, 1060 Vienna,  Austria        
\and  Department of Astronomy, University of Geneva, Ch. d’Ecogia 16, 1290 Versoix, Switzerland 
\and Kapteyn Astronomical Institute, University of Groningen, P.O. Box 800, 9700 AV Groningen, The Netherlands
\and Max-Planck-Institut für extraterrestrische Physik, Giessenbachstrasse 1, 85748 Garching, Germany
\and SUPA, School of Physics \& Astronomy, University of St. Andrews, North Haugh, St. Andrews KY16 9SS, UK
}

  \date{Received 20 Mar 2017 / Accepted 9 May 2017}

  \abstract
   {Episodic accretion is an important process in the evolution of young stars and their environment. The observed strong luminosity bursts of young stellar objects likely have a long lasting (i.e. longer than the burst duration) impact on the chemical evolution of the disk and envelope of young stars.}
   {We aim to investigate the observational signatures of the chemical evolution in the post-burst phase for embedded sources. With such signatures it is possible to identify targets that experienced a recent luminosity burst.}
   {We present a new model for the chemistry of episodic accretion based on the two dimensional, radiation thermo-chemical disk code P{\tiny RO}D{\tiny I}M{\tiny O} (PROtoplanetary DIsk MOdel). We have extended P{\tiny RO}D{\tiny I}M{\tiny O} with a proper treatment for envelope structures. For a representative Class~I model, we calculated the chemical abundances in the post-burst phase and produced synthetic observables such as intensity maps and radial intensity profiles.}
   {During a burst, many chemical species, such as CO, sublimate from the dust surfaces. As the burst ends they freeze out again (post-burst phase). This freeze-out happens from inside-out due to the radial density gradient in the disk and envelope structure. This inside-out freeze-out produces clear observational signatures in spectral line emission, such as rings and distinct features in the slope of radial intensity profiles. We fitted  synthetic \mbox{$\mathrm{C^{18}O}\:J\!=\!2\!-\!1$} observations with single and two component fits and find that post-burst images are much better matched by the latter. Comparing the quality of such fits therefore allows identification of post-burst targets in a model-independent way.}
   {Our models confirm that it is possible to identify post-burst objects from spatially resolved CO observations. However, to derive proper statistics, such as  the strength and frequencies of bursts, from observations it is important to consider the inclination and structure of the target and dust properties, as these have a significant impact on the freeze-out timescale.}

   \maketitle
%
%
\section{Introduction}
Protostellar accretion is an important constituent part of the star formation process, affecting the evolution of young stars and their circumstellar disks. It provides mass, angular momentum, and entropy for the nascent protostar and it feeds gravitational energy of accreted matter back to the disk via accretion luminosity, which, in the early evolution, can dominate the photospheric luminosity of the protostar \citep{Elbakyan2016}. Notwithstanding its importance, the character of protostellar accretion is poorly known, mainly due to the difficulty with observing the deeply embedded sources, and several theoretical and empirical models have been proposed to explain how young stars accumulate their mass \citep[e.g.][]{Shu1977,Hartmann1985,Bonnell1997,McKee2003c}.

Among these theories, the paradigm of variable accretion with episodic bursts \citep{Hartmann1985} has recently gained much attention thanks to ample indirect evidence  that protostellar accretion may have a highly variable character with prolonged episodes of low-rate (quiescent) accretion punctuated with short, but intense accretion bursts (see a review by \citealt{Audard2014}). Episodic accretion can have numerous and interesting implications not only for the disk dynamical and chemical evolution, but also for the evolution of pre-main-sequence stars. For instance, prolonged periods of low accretion luminosity between the bursts can promote disk gravitational fragmentation \citep{Stamatellos2012}. Episodic luminosity bursts can affect the disk and envelope chemical composition \citep{Lee2007cf,Visser2012m,Vorobyov2013f,Jorgensen2015b} and increase the growth rate of dust particles facilitating giant planet formation \citep{Hubbard2017}.

Finally, variable accretion with episodic bursts can help to resolve the `luminosity problem' of embedded protostars \citep{Dunham2012}, explain the existence of the very low luminosity objects (VELLOs) in the protostellar phase \citep{Vorobyov2017}, and affect the positions of pre-main-sequence stars on the HR diagram \citep{Baraffe2012,Hosokawa2011,Vorobyov2017a}. Until recently, episodic bursts were a feature exclusively attributed to low-mass star formation, but recent numerical models and observations demonstrated that massive stars can also have accretion bursts \citep{Burns2016,CarattioGaratti2016,Meyer2017}.

Two of the most pressing questions concerning episodic accretion are: do all young stars experience accretion bursts and what is the frequency of these bursts? The number of directly detected objects showing strong optical bursts remains low \citep[see e.g.][]{Audard2014}, as the typical timescale for the burst duration is about $100\,\mathrm{yr}$. It is therefore difficult to derive conclusive statistics and make a firm statement about the universality of the accretion burst phenomenon.

One option to identify outburst sources is through chemistry. During a burst the surrounding environment of young stars is heated up and molecules, frozen out on dust grains, can sublimate. One observational consequence of this scenario is the outward shift of ice lines. The ALMA (Atacama Large Millimeter Array) observations of the FU Ori source  \object{V883 Ori} indicate a location of the water ice line in the disk at $42\,\mathrm{au}$ \citep{Cieza2016}. This is significantly farther away than the expected location of the water ice line in disks of young low mass stars (typically one to five au).

After the burst the environment cools down quickly (less than one day, \citealt{Johnstone2013}) and the molecules can freeze out again \citep{Lee2007cf,Kim2011h,Kim2012a,Visser2012m,Vorobyov2013f,Visser2015c}. The freeze-out happens on a timescale of approximately $1000$ to $10^4\,\mathrm{yr}$, up to two order of magnitudes longer than a typical burst. Detections of objects currently in this post-burst phase, where the molecules are not yet frozen-out again, would significantly increase the statistical sample of known sources that experienced an episodic accretion event. 

There are already observational indications for chemistry driven by accretion bursts. \citet{Kim2011h,Kim2012a} argue that the high $\mathrm{CO_2}$ ice column densities measured in the envelopes of young stars can be explained by efficient conversion of CO to $\mathrm{CO_2}$ during burst-phases. However, the chemical details of such a process are still unclear. 

\citet{Jorgensen2013o} found clear indications of a recent burst in \object{IRAS 15398-3359}. Their spatially resolved ALMA observations show a lack of $\mathrm{HCO^+}$ close to the centre of the source although significant amounts of CO are detected. This $\mathrm{HCO^+}$ hole is likely produced by the efficient destruction of $\mathrm{HCO^+}$ by water that sublimated during a burst and has not yet frozen out again.

A further indication is the detection of extended \mbox{$\mathrm{C^{18}O}\:J\!=\!2\!-\!1$} emission in eight out of a sample of 16 Class 0/I sources observed with SMA (Submillimeter Array). By comparing the observations to 1D models of protostellar envelopes \citet{Jorgensen2015b} find that about half of their targets show extended emission with respect to their current bolometric luminosity. Again this can be explained by a recent burst and the delayed freeze-out of CO in the post-burst phase where the objects show their quiescent bolometric luminosity again. As discussed by \citet{Jorgensen2015b} only rough estimates concerning the burst frequency can be made from their sample as the sample size is still low and their results depend on the chosen binding energy of CO. A similar scenario is proposed by \citet{Kospal2016a} to explain the measured low degree of CO depletion in the disk of the EXOr prototype \object{EX Lupi}.

\citet{Visser2015c} modelled line fluxes of a diverse sample of Class 0/I objects with a combined 1D dust radiative transfer and sophisticated chemical model. They identified several line ratios to measure the time passed since the last accretion burst. However, the values derived from different line ratios show a large scatter. They concluded that one reason for this scatter might be their too-simple 1D structure model.

In order to put the chemical diagnostic of bursts on a more sophisticated footing, we introduce here a new two dimensional model for the chemistry of episodic accretion. This model is based on the radiation thermo-chemical disk code P{\tiny RO}D{\tiny I}M{\tiny O} (PROtoplanetary DIsk MOdel, \citealt{Woitke2009a,Kamp2010,Thi2011,Woitke2016}). We apply P{\tiny RO}D{\tiny I}M{\tiny O} to calculate the dust temperature, radiation field and the chemical abundances during the burst and in the post-burst phase. For a proper treatment of the remaining envelope of embedded sources we extended P{\tiny RO}D{\tiny I}M{\tiny O} with a parametric prescription for the envelope density structure. As a first application of this new 2D model we study the chemical evolution of gas-phase CO and the resulting observational signatures in the post-burst phase by means of synthetic observations of \mbox{$\mathrm{C^{18}O}\:J\!=\!2\!-\!1$} for a representative Class I model. We also investigate the impact of the disk component and inclination of the target on observables. We argue that the radial intensity profiles for \mbox{$\mathrm{C^{18}O}\:J\!=\!2\!-\!1$} show distinct signatures that allow for the identification of targets in the post-burst phase in a model independent way, in particular independent of the CO binding energy.

In Sect.~\ref{sec:method} we describe the physical structure of our model, the chemical network we use and how we simulate a burst. We discuss the chemical evolution of gas-phase CO and present synthetic observations for the \mbox{$\mathrm{C^{18}O}\:J\!=\!2\!-\!1$} spectral line emission for different structures (e.g. with or without a disk component) and inclinations in Sect.~\ref{sec:results}. In Sect.~\ref{sec:discussion} we outline a new method to identify targets in the post-burst phase and compare our results to other models. Finally we present our conclusions in Sect.~\ref{sec:conclusions}.
\section{Method}
\label{sec:method}
We modelled a Class~I burst scenario using the radiation thermo-chemical disk code P{\tiny RO}D{\tiny I}M{\tiny O} \citep{Woitke2009a,Kamp2010,Thi2011,Woitke2016}. We applied P{\tiny RO}D{\tiny I}M{\tiny O} to solve the wavelength dependent continuum radiative transfer which provides the temperature structure and the local radiation field for a given fixed density structure. On top of this we solved for the chemical abundances using time-dependent chemistry to follow the chemical evolution during the burst and post-burst phase. With the line transfer module of P{\tiny RO}D{\tiny I}M{\tiny O} \citep{Woitke2011} we produced synthetic observables (spectral line cubes) and used the CASA ALMA simulator to produce realistic images and radial intensity profiles.
\subsection{Physical model}
\subsubsection{Gas density structure}
\label{sec:struc}
The physical and structure parameters of our representative Class~I model are based on the model of \citet{Whitney2003}. The \citet{Whitney2003} model includes a disk component and an envelope structure with an outflow cavity. For the envelope geometry they use the infall solution including rotation \citep[e.g.][]{Ulrich1976,Terebey1984a} and for the disk a flared density structure is assumed.

Similar to \citet{Whitney2003} we calculated the two density structures independently and put the disk structure on top of the envelope structure wherever the disk density is higher than the envelope density. In the following we describe in detail the envelope and disk structure and the chosen parameter values. 

For the envelope structure we used the infalling rotating envelope model of \citet{Ulrich1976} where the density $\rho$ in units of $\mathrm{[g\,cm^{-3}]}$ is given by 
\begin{equation}
\rho(r,\mu)=\frac{\dot{M}_\mathrm{if}}{4\pi}
\left(2 G M_\mathrm{*}r^3\right)^{-1/2}
\left(\frac{1}{2}+\frac{\mu}{2\mu_0}\right)^{-1/2}
\left(\frac{\mu}{\mu_0}\!+\!\frac{2\mu_0^2 R_\mathrm{c}}{r}\right)^{-1}.
\label{eqn:envstruc3}
\end{equation}
$\dot{M}_\mathrm{if}$ is the mass infall rate of the envelope, $G$ the gravitational constant, $M_\mathrm{*}$ is the stellar mass, $r$ is the radial distance to the star in the centre, $R_\mathrm{c}$ is the centrifugal radius, $\mu=\cos\theta$ is the cosine polar angle of a streamline of infalling particles and $\mu_{0}$ is the value of $\mu$ far from the protostar ($r\rightarrow\infty$). The streamline angles are calculated via the equation
\begin{equation}
\mu_{0}^{3}+\mu_{0}(r/R_\mathrm{c}-1)-\mu(r/R_\mathrm{c})=0\label{eq:3}.
\end{equation}
Class~I objects show prominent bipolar cavities. To account for the cavity we again followed the approach of \citet{Whitney2003}. We used a cavity with an opening angle of $20\degr$ and subsequently applied a simple power-law to account for a curved cavity shape \citep{Whitney2003}. We assumed that the cavity is empty. The parameters for the envelope density structure are listed in Table~\ref{table:model}. For those parameters the resulting mass of the envelope is $M_\mathrm{env}\approx0.2\,\mathrm{M_\sun}$. 

For the disk component we used an axisymmetric flared gas density structure with a Gaussian vertical profile, and a power law with a tapered outer edge for the radial surface density profile \citep[e.g.][]{Lynden-Bell1974b,Andrews2009a}. The density structure for an azimuthally symmetric disk in hydrostatic equilibrium as a function of the spatial coordinates $r$ (distance to the star) and $z$ (height of the disk) is given by
\begin{equation}
  \label{eqn:density}
    \rho(r,z)=\frac{\Sigma(r)}{\sqrt{2\pi}\cdot h(r)}\exp\left(-\frac{z^2}{2h(r)^2}\right)\;\;\mathrm{[g\,cm^{-3}]}\;,   
\end{equation} 
where $\Sigma(r)$ describes the radial surface density profile and $h(r)$ is the scale height of the disk. For the surface density $\Sigma(r)$ we assumed a simple power-law distribution with a tapered outer edge
\begin{equation}
\Sigma(r)=\Sigma_0 \left(\frac{r}{R_\mathrm{in}}\right)^{-\epsilon}\exp\left(-\left(\frac{r}{R_{\mathrm{tap}}}\right)^{2-\gamma}\right)\;\;\mathrm{[g\,cm^{-2}]}\;, 
\label{eqn:surfdens}
\end{equation}
where $R_\mathrm{tap}$ is the characteristic radius and $R_\mathrm{in}$ the inner radius of the disk. The constant $\Sigma_0$ is determined via the relation $M_\mathrm{disk}=2\pi\int\Sigma(r)r\mathrm{dr}$ and Eq.~\ref{eqn:surfdens}. For the disk parameters chosen here $\Sigma_0=270\,\mathrm{[g\,cm^{-2}]}$. The vertical scale height $h(r)$ is described by a power law with a flaring power index $\beta$:
\begin{equation}
h(r)=H\mathrm{(100\;au)}\left(\frac{r}{100\;\mathrm{au}}\right)^{\beta}
\end{equation} 
where $H\mathrm{(100\;au)}$ gives the disk scale height at $r=100\,\mathrm{au}$.

The two density structures for the envelope and the disk were calculated independently. To merge the two structures we took the higher value of the density from the two components at a certain point of the domain. To secure a rather smooth transition at the outer border of the disk we varied the power index $\gamma$ for the tapered outer edge of the disk. This means that the disk structure in the outer regions no longer follows the viscous evolution (i.e. $\epsilon \ne \gamma$). 

For the inner radius of the structure we used $R_\mathrm{in}=0.6\,\mathrm{au}$ which corresponds roughly to the dust condensation radius during the burst (assuming that the burst happens close to the star, Sect.~\ref{sec:outburst}). The total hydrogen number density $n_\mathrm{<H>}=n_\mathrm{H}+2n_\mathrm{H_2}$ at the inner border is $n_\mathrm{<H>}(R_\mathrm{in})\approx2\times10^{14}\,\mathrm{cm^{-3}}$. The structure extends out to $R_\mathrm{out}=5000\,\mathrm{au}$ with $n_\mathrm{<H>}(R_\mathrm{out})\approx6\times10^{4}\,\mathrm{cm^{-3}}$. 

\begin{figure*}
\centering
\includegraphics{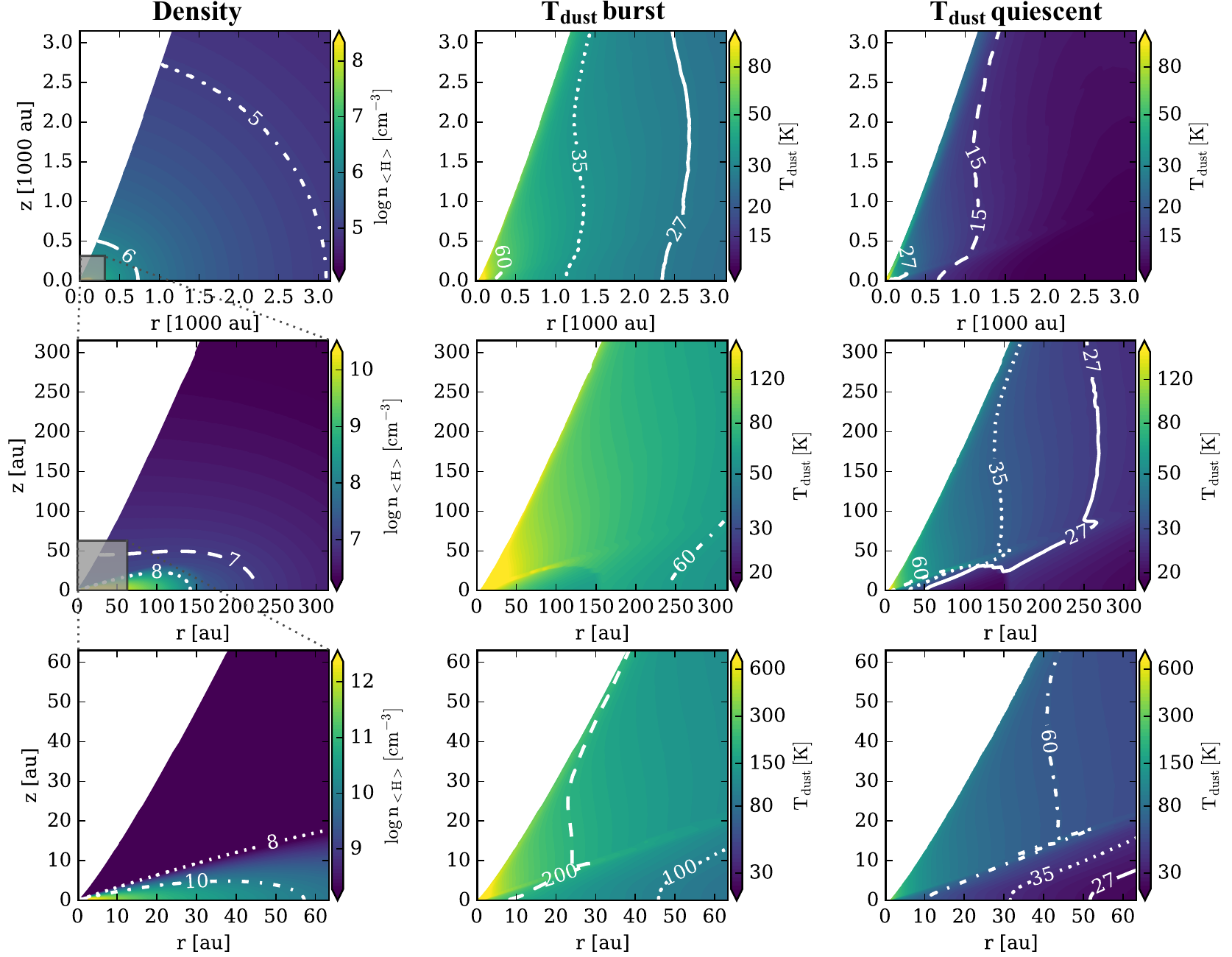}
\caption{Density and temperature structure for the representative Class~I model. From left to right: total hydrogen number density $n_\mathrm{<H>}$, $T_\mathrm{dust}$ during the burst, and $T_\mathrm{dust}$ in the quiescent (post-burst) phase. The second and third row show a zoom-in to the inner $300$ and $60~\mathrm{au}$ (marked by the grey squares in the density plots). The white solid contour in the temperature plots for $T=27\,\mathrm{K}$ roughly indicates the CO ice line. Some artefacts are visible in the temperature structure (e.g. last two panels in the second row). These are due to the sharp transition of the two dust populations used for the disk and envelope, but have no significant impact on the large scale temperature structure.}
\label{fig:structure}
\end{figure*}
All parameters of the model are listed in Table~\ref{table:model}. The resulting gas density structure of our representative Class~I model is shown in the first column of Fig.~\ref{fig:structure}. 
\begin{table}
\caption{Main parameters of the Class I model.}
\label{table:model}
\centering
\begin{tabular}{l|c|c}
\hline\hline
Quantity & Symbol & Value\\
\hline
stellar mass                          & $M_\mathrm{*}$                    & $0.5~\mathrm{M_{\sun}}$\\
stellar effective temp.               & $T_{\mathrm{*}}$                  & 5000~K\\
stellar luminosity                    & $L_{\mathrm{*}}$                  & $1.0~\mathrm{L_{\sun}}$\\
\hline
disk gas mass                         & $M_{\mathrm{disk}}$               & $0.02~\mathrm{M_{\sun}}$\\
disk inner radius                     & $R_{\mathrm{in}}$                 & 0.6~au\\
disk tapering-off radius              & $R_{\mathrm{tap}}$                & 100~au\\
column dens. pow. ind.              & $\epsilon$                        & 1.0\\
tapering parameter                    & $\gamma$                          & -1.0\\
reference scale height                & $H(100\;\mathrm{au})$             & 10 au\\
flaring power index                   & $\beta$                           & 1.1\\
\hline
centrifugal radius                    & $R_{\mathrm{c}}$                & 100~au\\
mass infall rate                      & $\dot{M}_\mathrm{if}$                          & $5\times10^{-6}~\mathrm{M_\sun\,yr^{-1}}$\\   
outer radius                          & $R_{\mathrm{out}}$                 & 5000~au\\
cavity opening angle                  & $\beta_\mathrm{cav}$               & $20^{\circ}$ \\
\hline
dust to gas mass ratio                & $\delta$                              & 0.01 \\
min. dust particle radius             & $a_\mathrm{min}$                  & $\mathrm{0.005~\mu m}$\\
max. dust particle radius             & $a_\mathrm{max}$                  & \\
~~envelope                                &                                   & $\mathrm{1~\mu m}$\\
~~disk                            &                                   & $\mathrm{1000~\mu m}$\\
dust size dist. power ind.           & $a_\mathrm{pow}$                  & 3.5\\
dust composition\tablefootmark{a}    & {\tiny Mg$_{0.7}$Fe$_{0.3}$SiO$_3$} & 60\%\\
(volume fractions)                    & {\tiny amorph. carbon}           & 20\%\\
                                      & {\small vacuum}                   & 20\%\\
\hline
cosmic ray $\mathrm{H_2}$ ion. rate & $\mathrm{\zeta_{CR}}$               &$\mathrm{5\times10^{-17}s^{-1}}$\\
strength of interst. FUV              & $\chi^\mathrm{ISM}$               & 1\tablefootmark{b}\\
\hline
distance                              &  $d$                                & 200 pc\\
\hline
\end{tabular}
\tablefoot{
\tablefoottext{a}{Optical constants are from \citet{Dorschner1995a} and \citet[][BE-sample]{Zubko1996c}; resulting dust material density $\rho_\mathrm{dp}\approx2.2\,\mathrm{g\:cm^{-3}}$.}
\tablefoottext{b}{$\chi^\mathrm{ISM}$ is given in units of the Draine field \citep{Draine1996b,Woitke2009a}.}
}
\end{table}
\subsubsection{Dust properties}
\label{sec:dustprop}
We assumed a canonical dust to gas mass ratio of 0.01 for the disk and envelope structure. Observations and models show clear evidence for dust growth in disks \citep[e.g.][]{Birnstiel2010a,Williams2011,Pinte2016}. We accounted for this by using two different dust populations for the disk and the envelope structure. 

For both structures we used a simple power law for the dust size distribution function $f(a)\propto a^{-a_\mathrm{pow}}$, where $a$ is the grain radius. In the disk the minimum and maximum dust sizes are $a_\mathrm{min}=0.005\,\mathrm{\mu m}$ and $a_\mathrm{max}=1000\,\mathrm{\mu m}$, respectively. Scattered light images from dense cloud cores (`coreshine') indicate that the maximum dust particle size is likely larger than in the diffuse interstellar medium \citep{Steinacker2015b,Ysard2016}. We therefore adopted a value of $a_\mathrm{max}=1\,\mathrm{\mu m}$ for the envelope. For both dust size distributions we used 300 logarithmically spaced size bins. For the slope we adopted the canonical value for interstellar grains of $a_\mathrm{pow}=3.5$ \citep{Mathis1977b}.

\begin{figure}
\includegraphics{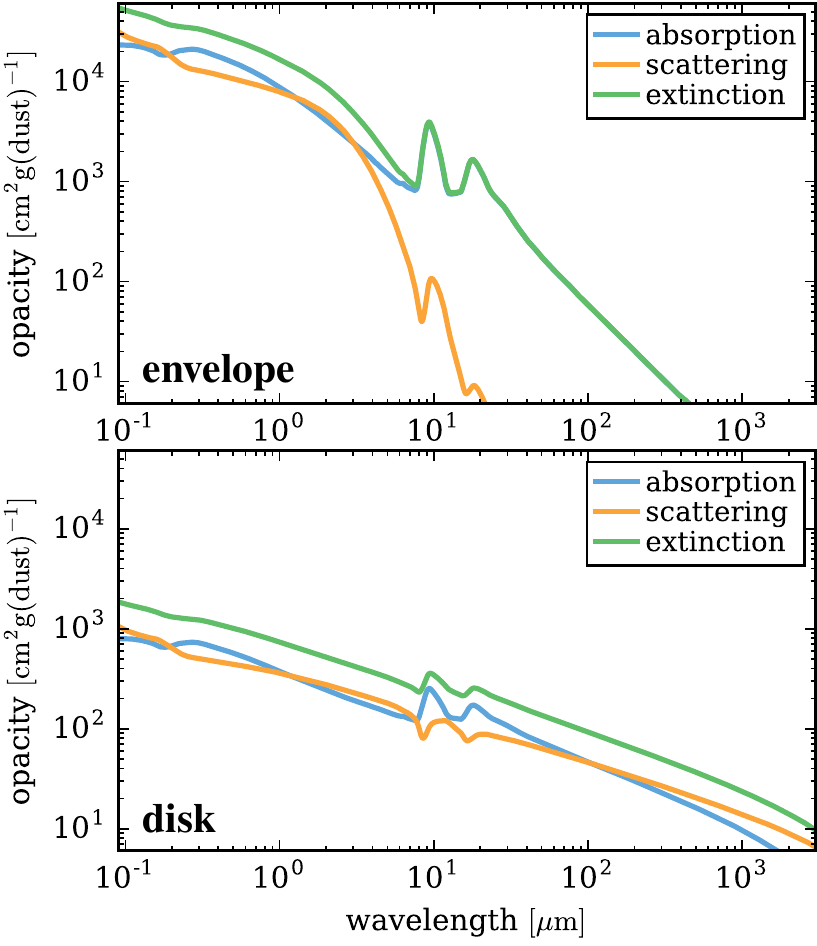}
\caption{Dust opacities for the envelope and the disk dust population.}
\label{fig:dustopac}
\end{figure}

For the dust composition we assumed a mixture of 60\% (by volume) amorphous laboratory silicate, 20\% amorphous carbon and 20\% vacuum (porosity of grains). These values are similar to the proposed dust composition of \citet{Woitke2016} for modelling of T~Tauri disks. To calculate the dust opacities we applied Mie theory \citep{Mie1908}. The detailed dust properties are given in Table~\ref{table:model}, the resulting dust opacities for both dust populations are shown in Fig.~\ref{fig:dustopac}.

We did not include ice-coated grains in our opacity calculations. Such grains would show a distinct opacity feature at $\approx3\,\mathrm{\mu m}$ \citep[e.g.][]{Ossenkopf1994a}. We also ignored any change in the dust opacity which might be caused by the burst itself, such as the observed crystallization of dust particles in the disk surface layers of \object{EX Lupi} \citep{Abraham2009}. Although such opacity features are clearly observable in spectral energy distributions they do not have a significant impact on our results. This is because we are only interested in the overall change in the dust temperature structure during a burst and do not model spectral energy distributions in detail (see also Appendix~\ref{sec:visser}).
\subsection{Chemical model}
\label{sec:chemmodel}
Our chemical network is based on the UMIST~2012 database for gas phase chemistry \citep{McElroy2013b}. We used a reduced network and only selected chemical reactions from UMIST~2012 for a set of 76 gas phase species. We expanded the gas phase network by reactions for adsorption and desorption for all neutral species (excluding atomic and molecular hydrogen and noble gases). We also considered H$_2$ formation on dust grains \citep{Cazaux2002,Woitke2009a}. For the adsorption and desorption reactions we used the binding energies from the UMIST~2012 release but updated a couple of values (see Table~\ref{tab:adsenergies}) to be consistent with \citet{Visser2015c}. In total our network consists of 105 species and 1211 chemical reactions. For more details on the chemical network see also \citet{Kamp2017}.

We used time-dependent chemistry to follow the evolution of the chemical abundances during and after the burst (see Sect.~\ref{sec:outburst}). As initial conditions for time-dependent chemistry we chose the same abundances as \citet{Visser2015c}, which are typical for prestellar core conditions (see Table~\ref{tab:tcheminit}). 

The most important chemical process for episodic accretion chemistry is the freeze-out and sublimation (adsorption and desorption) of neutral species \citep[e.g.][]{Visser2015c}. Besides thermal desorption we also include non-thermal desorption by cosmic rays and photons \citep{Woitke2009a,Kamp2017}. However, the most relevant aspect of episodic accretion chemistry is the balance between adsorption and thermal desorption. Therefore we explain the modelling of these processes in P{\tiny RO}D{\tiny I}M{\tiny O} in detail.
\subsubsection{Adsorption}
\label{sec:adsorption}
For the chemistry we did not use a fixed single dust size but derived averaged dust sizes from the same dust size distributions as were used for the dust opacity calculations \citep{Woitke2009a,Woitke2016}. The adsorption (freeze-out) rate per volume for a species $i$ is given by 
\begin{equation}
\label{eq:adsorption}
R_{i\mathrm{,ads}}=n_i4\pi\langle a^2\rangle n_\mathrm{d}\alpha v_{i,\mathrm{th}}\;\;\mathrm{[cm^{-3}s^{-1}}]
\end{equation}
where $\langle a^2\rangle=\int_{a_\mathrm{min}}^{a_\mathrm{max}}a^2f(a)\mathrm{d}a$ is the second moment of the dust size distribution, $n_\mathrm{d}$ is the dust particle number density and $\alpha=1$ the sticking efficiency. The thermal velocity for species $i$ is given by $v_{i,\mathrm{th}}=\sqrt{kT_\mathrm{gas}/2\pi m_i}$, where $T_\mathrm{gas}$ is the gas temperature and $m_i$ the mass of the species in [g].
\begin{table}
\caption{Dust properties of relevance for the chemistry in the disk, envelope and for a uniform distribution with $a=0.1\,\mathrm{\mu m}$.} 
\label{tab:dustprop}
\resizebox{\hsize}{!}{
\begin{tabular}{lccc}
\hline\hline
 & disk & envelope & $0.1\,\mathrm{\mu m}$ \\
\hline
Mean dust size $\langle a^2\rangle^{1/2}\,\mathrm{[\mu m]}$& $0.0112$ & $0.0108$ & $0.1$\\
Mean dust size $\langle a^3\rangle^{1/3}\,\mathrm{[\mu m]}$& $0.0653$ & $0.0202$ & $0.1$\\
Particle abundance $n_d/n_\mathrm{<H>}$ & $9.0(-12)\tablefootmark{a}$ & $3.0(-10)$ & $2.5(-12)$ \\
Surface $4\pi\langle a^2\rangle n_\mathrm{d}/n_\mathrm{<H>}\,\mathrm{[cm^{2}]}$ & $1.4(-22)$ & $4.4(-21)$ & $3.2(-21)$  \\
\hline
\end{tabular}}
\tablefoot{Here we assumed the same dust composition as given in Table~\ref{table:model} for the disk, envelope and the $0.1\,\mathrm{\mu m}$ cases.
\tablefoottext{a}{$x(y)$ means $x\times10^{y}$.}}
\end{table}
In Table~\ref{tab:dustprop} several important dust size distribution quantities for the chemistry are listed. For comparison we also list the values assuming only a single dust size of $0.1\,\mathrm{\mu m}$ (see also \citealt{Woitke2016}). The adsorption rate is the relevant quantity for the freeze-out timescale. We will discuss this in detail in Sect.~\ref{sec:freezetimescale}.
\subsubsection{Thermal desorption}
The thermal desorption rate of a species $i$ is given by  
\begin{equation}
\label{eq:desorption}
R_{i\mathrm{,des}}=n_{i\mathrm{\#,des}}\nu_{i\mathrm{,osc}}\exp\left(-\frac{E_{i\mathrm{,B}}}{kT_\mathrm{d}}\right)\;\;\mathrm{[cm^{-3}s^{-1}}]. 
\end{equation}
$n_{i\mathrm{\#,des}}$ is the fraction of the number density of species $i$ on the dust grains (\# indicates the ice phase) prone to desorption (see below). $\nu_{i\mathrm{,osc}}=\sqrt{(2n_\mathrm{surf}kE_{i\mathrm{,B}}/(\pi^2m_i))}$ is the vibrational frequency where $n_\mathrm{surf}=1.5\times10^{15}\,\mathrm{cm^{-2}}$ \citep{Hasegawa1992au} is the surface density of available adsorption sites and $E_{i\mathrm{,B}}$ is the adsorption binding energy for species $i$. The vibrational frequency for CO is $\nu_\mathrm{osc}(\mathrm{CO})=1.1\times10^{12}\,\mathrm{s^{-1}}$ for $E_\mathrm{B}(\mathrm{CO})=1307\,\mathrm{K}$.

We assumed that only a certain fraction of the total (thick) ice mantle on the dust grain is effectively desorbed \citep{Aikawa1996h,Aikawa2015}. The total number density of active surface places (i.e. prone to desorption) in the ice mantle is given by
\begin{equation}
n_\mathrm{act\#}=4\pi\langle a^2\rangle n_\mathrm{d}n_\mathrm{surf}N_\mathrm{lay}, 
\end{equation} 
where $\langle a^2\rangle$ is the second moment of the dust size distribution (see Sect.~\ref{sec:adsorption}), $n_\mathrm{d}$ the number density of dust particles and $N_\mathrm{lay}$ is the number of active layers. We adopted $N_\mathrm{lay}=2$ (i.e. only the outermost two layers can be desorbed, \citealt{Aikawa1996h}). The number density $n_{i\mathrm{\#,des}}$ of active ice units for a species $i$ is given by \citep{Woitke2009a} 
\begin{equation}
n_{i\mathrm{\#,des}}=\begin{cases}
    n_{i\mathrm{\#}}, & \text{if $n_\mathrm{tot\#}<n_\mathrm{act\#}$}.\\
    n_\mathrm{act\#}\frac{n_{i\mathrm{\#}}}{n_\mathrm{tot\#}}, & \text{otherwise}.
  \end{cases}
\end{equation}
where $n_{i\mathrm{\#}}$ is the number density of the ice species $i$ and $n_\mathrm{tot\#}$ the sum of the number density of all ice species. In the case of thick ice mantels ($n_\mathrm{tot\#} \geq n_\mathrm{act\#}$) the desorption rate Eq.~\ref{eq:desorption} is of zeroth order, which means that the $R_\mathrm{des}$ does not (strongly) depend on the actual number density of the considered ice species \citep[e.g.][]{Collings2004d,Collings2015}. In case of thin ice mantles (i.e. not all available active adsorption sites are occupied), Eq.~\ref{eq:desorption} transforms to a first-order desorption rate, which means that $R_\mathrm{i,des}$ scales linearly with $n_\mathrm{i\#}$. 

According to the temperature-programmed desorption (TPD) laboratory experiments zero-order desorption should be the preferred method to estimate desorption rates \citep{Collings2004d,Collings2015}. For our method to estimate the thermal desorption rate, actually in most areas of our model structure zero-order desorption applies (i.e. roughly speaking everywhere where water is frozen-out). The main effective difference between first-order and zero-order desorption is that in the case of zero-order desorption the number of ice molecules which can sublimate is limited. As a consequence the residual gas phase abundances in the freeze-out zone is lower for zero-order desorption compared to first-order desorption where all molecules in the ice mantle are prone to desorption. 

In Appendix~\ref{sec:visser} we compare our chemical model to the model of \citet{Visser2015c}. We find a good agreement between the two models in particular concerning the main aspects of episodic accretion chemistry (e.g. delayed freeze-out, shift of ice-lines). 
\subsection{Burst scenario}
\label{sec:outburst}
To simulate a luminosity burst we followed the approach of \citet{Visser2015c}. For the burst they assume an instantaneous increase of the protostellar luminosity by a factor of 100, compared to the quiescent phase, and a burst duration of $100\,\mathrm{yr}$. After the burst the luminosity drops instantaneously to the quiescent value. The envelope density structure is kept fixed at all times (i.e. any dynamical changes of the structure are neglected, see also Sect.~\ref{sec:dynamicalevol}).

Just increasing the stellar luminosity is a very simplified picture of an episodic accretion event. Most of the excess energy during the burst is actually produced in a small and hot accretion disk close to the protostar ($r<1\,\mathrm{au}$; \citealt{Zhu2007jw,Zhu2008}). However, for the chemistry mainly the resulting temperature change in the envelope and disk structure matters and the process producing the excess luminosity during the burst is not relevant. Furthermore, the inner region of the surrounding dust structure is optically thick and the photons emitted by the protostar and accretion disk are reprocessed to longer wavelengths. Consequently, the temperature in regions farther out is not sensitive to the details of the inner region as the outer regions mainly sees the reprocessed photons \citep{Johnstone2013}. 

To model the emission of the central luminosity source we used PHOENIX stellar atmosphere models \citep{Brott2005a} for a given stellar mass $M_\mathrm{*}=0.5\,\mathrm{M_\sun}$, luminosity $L_\mathrm{*}=1.0\,\mathrm{L_\sun}$ and effective temperature $T_\mathrm{*}=5000\,\mathrm{K}$. For the burst we increased $L_\mathrm{*}$ by a factor of 100 but kept all other stellar parameters fixed. We note those parameters should not be interpreted as proper protostellar parameters but rather as a very simple approximation to simulate the energy output produced close to the protostar during a burst. We performed some tests varying the properties of the input stellar spectrum (e.g. $T_\mathrm{*}$) and also used the burst spectrum of \object{FU Orionis} \citep{Zhu2007jw} as input. We find that our results presented here are not strongly sensitive to the shape of the used stellar spectrum. The increase of the dust temperature during the burst is mostly proportional to the luminosity ($T_\mathrm{d}\propto L_\mathrm{*}^{0.25}$; see \citealt{Johnstone2013} and Appendix~\ref{sec:visser}).

In the following we describe the main steps of our model to simulate the chemical evolution in the post-burst phase.
\begin{enumerate}
 \item\emph{quiescent RT:} continuum radiative transfer using the quiescent protostellar luminosity to calculate the local radiation field and temperatures used for the chemical evolution in the pre-burst and post-burst phases.   
\item\emph{init chemistry:} evolve the chemistry, starting with prestellar core abundances (see Table~\ref{tab:tcheminit}), under quiescent conditions (i.e. temperature structure) for $10^5\,\mathrm{yr}$ to calculate the initial chemical abundances prior to the burst. We will discuss the impact of the initial chemical abundances on our results in Sect.~\ref{sec:initcond}; 
  \item\emph{burst RT:} continuum radiative transfer (RT) for the given burst luminosity to calculate the local radiation field and temperatures at every point of the structure; 
  \item\emph{burst chemistry:} evolve the chemistry for the duration of the burst starting from given initial abundances. During the chemistry step the radiation field and temperatures are kept fixed at the burst values; 
  \item\emph{post-burst chemistry:} follow the chemical evolution for $10^5\,\mathrm{yr}$ under quiescent conditions and produce synthetic observations at distinct time steps.
\end{enumerate}
We have assumed here that the temperature change due to the burst and after the burst happens instantaneously. \citet{Johnstone2013} indeed find that the dust in a typical protostellar envelope very quickly reaches its equilibrium temperature after a luminosity change ($<1~\mathrm{day}$). We applied their semi-analytic method to our structure, including the disk, and find that the heating times can increase by at most two orders of magnitude. We also neglected any possible differences between the gas and dust temperature (see \citealt{Johnstone2013} for a discussion) and assume that the gas and dust temperatures are equal at all times. Considering the long timescales for the chemistry (see Sect.~\ref{sec:freezetimescale}), it is a reasonable simplification to assume that the temperature reacts instantaneously to the luminosity change of the central heating source (see also \citealt{Visser2015c}). 

The temperature structure for the burst and quiescent phases are shown in Fig.~\ref{fig:structure}. The main difference in the envelope temperature structure compared to 1D spherical models are the lower temperatures close to the midplane. The disk absorbs most of the stellar radiation and therefore casts a shadow onto the envelope, consequently the temperatures are lower within the disk shadow. This becomes also apparent by the temperature contours shown in Fig.~\ref{fig:structure} which are not circular.
\subsection{Synthetic observations}
\label{sec:syntheticobs}
To produce synthetic observations we use the built-in line radiative transfer module of P{\tiny RO}D{\tiny I}M{\tiny O} \citep{Woitke2011}. For the line radiative transfer we assume a Keplerian velocity field for the disk component and free-fall velocity for the envelope structure (i.e. neglect rotation of the envelope for simplicity).

To produce realistic images and radial intensity profiles for spectral line emission we convolve the line cubes produced by P{\tiny RO}D{\tiny I}M{\tiny O} with a given synthetic beam but also perform full ALMA/CASA simulations (see Appendix~\ref{sec:alma_method}). In this work we focus on the \mbox{$\mathrm{C^{18}O}\:J\!=\!2\!-\!1$} line. Our chemical model does not include CO isotopologue chemistry (e.g. \citealt{Visser2009s}). Instead, we calculated the abundance of $\mathrm{C^{18}O}$ by applying a fixed $\mathrm{^{16}O/^{18}O}$ isotopologue ratio of $498.7$ \citep[e.g.][]{Scott2006}. For the excitation calculations we used the collision rate coefficients for $\mathrm{C^{18}O}$ with $\mathrm{H_2}$ from \citet{Yang2010} provided by the LAMDA database (Leiden Atomic and Molecular Database, \citealt{Schoeier2005c}).
\section{Results}
\label{sec:results}
\subsection{Adsorption timescale}
\label{sec:freezetimescale}
The adsorption or freeze-out timescale is the most relevant quantity in episodic accretion chemistry. It defines the time range in which one can still see chemical signatures initially caused by an accretion burst.

At first we bring Eq.~\ref{eq:adsorption} (without $n_i$) in the same form as \citet{Charnley2001d}
\begin{equation}
\label{eq:adscharnley}
k_{i,\mathrm{ads}}=1.45\times10^{4}\cdot\alpha\cdot \left(\frac{T_\mathrm{gas}}{M_i}\right)^{1/2}\pi\langle a^2\rangle n_\mathrm{d}\,\,\mathrm{[s^{-1}]}, 
\end{equation}
where $M_i$ is the molecular weight of species $i$. In P{\tiny RO}D{\tiny I}M{\tiny O} the number density of dust particles $n_\mathrm{d}$ is given by 
\begin{equation}
\label{eq:ndust}
n_\mathrm{d}=\frac{\rho_\mathrm{gas}\cdot \delta}{\frac{4\pi}{3}\cdot \langle a^3\rangle \rho_\mathrm{dp}}\,\,\mathrm{[cm^{-3}]}, 
\end{equation}
where $\rho_\mathrm{gas}=2.28\times10^{-24}n_\mathrm{<H>}$ is the gas density in $\mathrm{[g\,cm^{-3}]}$, $\delta$ is the dust to gas mass ratio, $\langle a^3\rangle$ is the third moment of the dust size distribution and $\rho_\mathrm{dp}$ is the material density of a dust particle in $\mathrm{[g\,cm^{-3}]}$. The adsorption timescale $t_{i,\mathrm{ads}}={k_{i,\mathrm{ads}}}^{-1}$ can than be written in the form
\begin{equation}
\label{eq:freezetimescale}
t_\mathrm{i,ads}=2.9\times10^{-12}\cdot M_i^{1/2}\cdot\alpha^{-1}\cdot\frac{\rho_\mathrm{dp}}{\delta}\cdot T_\mathrm{gas}^{-1/2}\cdot\frac{\langle a^3 \rangle}{\langle a^2 \rangle}\cdot \rho_\mathrm{gas}^{-1}\;\;[\mathrm{yr}].   
\end{equation}
Equation~\ref{eq:freezetimescale} includes all the quantities and parameters of our model that have an impact on $t_\mathrm{i,ads}$. However, for the results presented here we do not vary all parameters, in particular $\alpha=1$ (sticking efficiency), $\rho_\mathrm{dp}\approx2.2\,\mathrm{g\,cm^{-3}}$ and $\delta=0.01$ are the same for all models and are constant in time and space.

Here, we want to focus on the impact of the dust grain size. The ratio $\langle a^3 \rangle/\langle a^2 \rangle$ represents a mean particle radius $\overline{a}$, where the averaging is surface weighted. This means $\overline{a}$ is actually dominated by the small dust particle population. Using $\overline{a}$ instead of a simple averaging over grain sizes is more appropriate to model the adsorption process, as the total available dust surface is most relevant here and the total surface is mainly provided by the large number of small grains (see also \citealt{Vasyunin2011}).

We use here two different dust populations for the disk and envelope, consequently also $\overline{a}$ varies. For our chosen dust populations $\overline{a}_\mathrm{disk}=2.24\,\mathrm{\mu m}$ and $\overline{a}_\mathrm{env}=0.07\,\mathrm{\mu m}$. This means that $t_\mathrm{ads}$ increases by a factor of $\approx32$ in the disk, compared to the case with only small dust grains (neglecting any temperature change).

In P{\tiny RO}D{\tiny I}M{\tiny O} the same dust properties are consistently used for the dust radiative transfer and the chemistry. In other chemical models usually a mean dust size and a dust abundance is assumed. Typical values are $\overline{a}=0.1\,\mathrm{\mu m}$ and $n_\mathrm{d}=10^{-12}n_\mathrm{<H>}$ which can directly be used in Eq.~(\ref{eq:adscharnley}) to calculate $t_{i,\mathrm{ads}}$ (e.g. Eq.~(3) of \citealt{Rodgers2003}). For such values $t_\mathrm{i,ads}$ is about a factor of three longer compared to our dust model for the envelope (for a given density and temperature). These examples show the importance of dust properties like the assumed material density and dust size for the adsorption timescale as already briefly discussed by \citet{Vorobyov2013f} (see also Sect.~\ref{sec:lumrad})

Our results concerning the $t_\mathrm{i,ads}$ show for the first time the importance of dust properties and evolution for the chemistry of episodic accretion in a quantitative manner. This indicates that in particular for more evolved sources dust evolution processes like dust growth but also radial migration and dust settling which are important especially for the disk \citep[see e.g.][]{Vasyunin2011,Akimkin2013}, should be taken into account for modelling the chemistry of episodic accretion.
\begin{figure*}
\centering
  \includegraphics{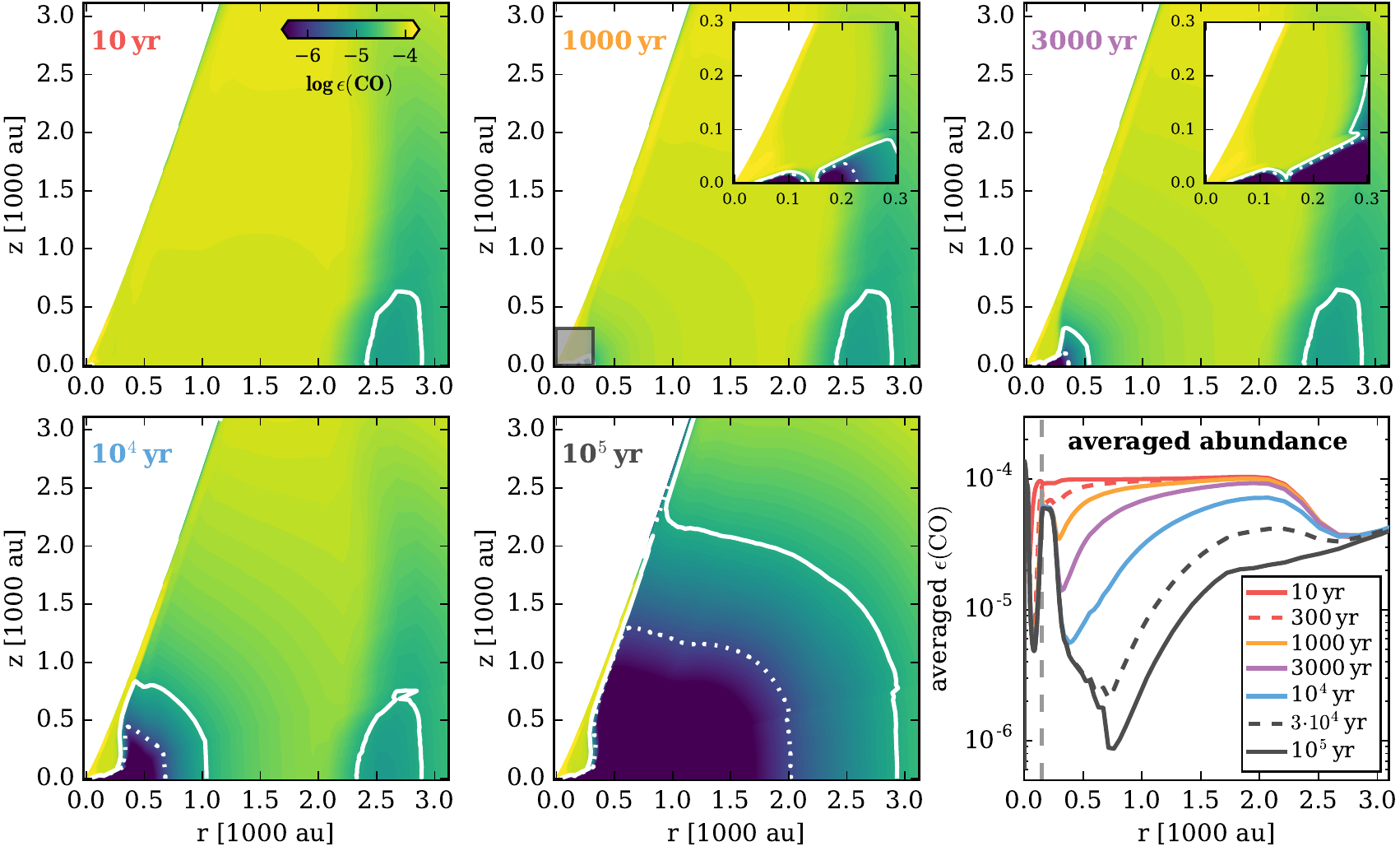}
  \caption{Evolution of the CO gas phase abundance $\mathrm{\epsilon(CO)}$ in the post-burst phase. The 2D contour plots show $\mathrm{\epsilon(CO)}$ for five different times of $t=10$, $1000$, $3000$, $10^4$ and $10^5\,\mathrm{yr}$ after the end of the burst. The white solid and dashed contours indicate $\mathrm{\epsilon(CO)}=10^{-5}$ and $\mathrm{\epsilon(CO)}=10^{-6}$, respectively. The $1000$ and $3000\,\mathrm{yr}$ panels include a zoom in for the inner $300~\mathrm{au}$ (marked by the grey box). The bottom right panel shows the evolution of the vertically averaged CO abundance as a function of the midplane radius (the dashed lines are for $t=300$ and $t=3\times10^4\,\mathrm{yr}$). The vertical grey dashed line marks the radial disk to envelope transition at $r\approx150\,\mathrm{au}$. The black solid line corresponds to the quiescent phase ($t=10^5\,\mathrm{yr}$).}
  \label{fig:CI_COabun}
\end{figure*}
\subsection{CO gas phase abundance}
\label{sec:COabundance}
In this section we present the CO gas phase abundance structure of our 2D model for the quiescent phase (i.e. no burst) and the detailed time evolution of the abundance in the post-burst phase. In the post-burst phase the envelope and disk have already cooled down to quiescent conditions (see Sect.~\ref{sec:outburst}) and CO sublimated during the burst can freeze out again.

In Fig.~\ref{fig:CI_COabun} we show the CO gas phase abundance $\epsilon(\mathrm{CO})=n_\mathrm{CO}/n_\mathrm{<H>}$ for the inner $3000\,\mathrm{au}$ of the 2D structure at five different times after the end of the burst. The bottom right panel in Fig.~\ref{fig:CI_COabun} shows the radial profiles of the vertically averaged CO abundance $\epsilon_\mathrm{avg}(\mathrm{CO})$. $\epsilon_\mathrm{avg}(\mathrm{CO})$, as a function of radius, is given by the ratio $\epsilon_\mathrm{avg}(\mathrm{CO})=N_\mathrm{CO,ver}/N_\mathrm{<H>,ver}$, where $N_\mathrm{CO,ver}$ is the vertical column density of gas-phase CO and $N_\mathrm{<H>,ver}$ is the total hydrogen column density $(N_\mathrm{<H,ver>}=N_\mathrm{H,ver}+2N_\mathrm{H_2,ver}$). At first we discuss the CO abundance pattern for the quiescent phase and subsequently the detailed evolution of the CO gas phase abundance shortly after the burst (post-burst phase).
\subsubsection{Quiescent phase}
Due to our choice for the initial chemical abundances prior to the burst (Sect.~\ref{sec:outburst}) the pre-burst CO abundance structure is identical to what is shown in the $t=10^5\,\mathrm{yr}$ panel in Fig.~\ref{fig:CI_COabun}, which we call the quiescent phase. We will discuss the consequences of the chosen initial abundances and the possible impact of recurrent bursts in Sect.~\ref{sec:initcond}. The main features of the averaged CO abundance profile in the quiescent phase are
\begin{itemize}
\item an inner region within the quiescent radial CO ice line ($r\approx300\,\mathrm{au}$) with an average CO abundance  close to the canonical value of \mbox{$\epsilon_\mathrm{avg}(\mathrm{CO})\approx2\times10^{-4}$} (the impact of the disk component is discussed in the following paragraphs);  
\item strong freeze-out of CO just outside the radial CO ice line with $\epsilon_\mathrm{avg}(\mathrm{CO})\lesssim{10^{-6}}$;
\item a gradual increase of $\epsilon_\mathrm{avg}(\mathrm{CO})$ with radius until  
$\epsilon_\mathrm{avg}(\mathrm{CO})$ reaches again the canonical value at the outskirts of the structure.
\end{itemize}
Such a quiescent profile is consistent with observations of embedded sources \citep[e.g.][]{Jorgensen2005,Yildiz2010,Anderl2016}, which in particular show the gradual increase of the CO abundance with radius, beyond the CO ice line. This implies that our simple model (i.e. assuming a steady-state structure) indeed captures the main characteristics of the CO abundance structure of embedded sources. 

In our model the detailed appearance of the positive slope of $\epsilon_\mathrm{avg}\mathrm{(CO)}$ for $r \gtrsim 750 \,\mathrm{au}$ has two reasons. Averaging the CO abundance vertically is not the ideal representation at large scales where the density structure is rather spherical. The second reason is of physical nature. Due to the lower densities the \mbox{freeze-out} becomes less efficient as collisions of molecules with dust particles become less likely. As a consequence non-thermal desorption processes such as cosmic-ray and photo desorption become, relatively speaking, more important. In addition the outer parts of the envelope ($r\gtrsim 1500\,\mathrm{au}$) are also affected by the interstellar background radiation field included in our model, further increasing the impact of photo-desorption in the outskirts of the envelope.
\subsubsection{Post-burst phase}
During the burst all CO sublimates in the inner $2000-3000\,\mathrm{au}$. Only there the temperature increases above the CO sublimation temperature of $T_\mathrm{sub}(\mathrm{CO})\approx27\,\mathrm{K}$ (see~Fig.~\ref{fig:structure}).  For $r\gtrsim2000-3000\,\mathrm{au}$ $\epsilon(\mathrm{CO})$ is not affected by the burst and the pre-burst abundances are preserved. In the first panel of Fig.~\ref{fig:CI_COabun} ($t=10\,\mathrm{yr}$), the impact of the disk on the post-burst CO abundance in the outer regions is also apparent. Around the midplane of the structure ($z=0\,\mathrm{au}$) the disk absorbs most of the stellar radiation and casts a shadow into the envelope. As a consequence CO sublimates only up to $r=2000\,\mathrm{au}$ in the shadowed region.
  
Looking at the averaged abundance panel in Fig.~\ref{fig:CI_COabun}, the evolution of $\epsilon_\mathrm{avg}(\mathrm{CO})$ in the post-burst phase shows three  distinct features:
\begin{itemize}
  \item the fast depletion of CO in the zone with the disk \mbox{($r\lesssim 150\,\mathrm{au}$)};
  \item the peak in $\epsilon_\mathrm{avg}(\mathrm{CO})$ at $150 \lesssim r \lesssim 300\,\mathrm{au}$; 
  \item the slow depletion of CO with time in the region \mbox{$300 \lesssim r \lesssim 2500\,\mathrm{au}$};
\end{itemize}     
In the zone with the disk, $\epsilon_\mathrm{avg}(\mathrm{CO})$ is mainly determined  by the high density in the disk. The temperatures close to the midplane of the disk are below $T_\mathrm{sub}(\mathrm{CO})$ and CO freezes out on a timescale of $\lesssim\!100\,\mathrm{yr}$. Only within the radial CO ice line (at $r\approx 50\,\mathrm{au}$ in the disk midplane) CO remains in the gas phase at all times. Above the disk (see the inset in the plot for $t=1000\,\mathrm{yr})$ the temperature is higher than $T_\mathrm{sub}(\mathrm{CO})$ and CO remains in the gas phase at all times. 

In the zone with $150 \lesssim r \lesssim 300\,\mathrm{au}$ CO is mostly in the gas phase but shows some depletion with $\epsilon_\mathrm{avg}(\mathrm{CO})\approx6\times10^{-5}$. The radius $r\approx 300\,\mathrm{au}$ can be seen as the radial CO ice line in the envelope (i.e. similar to a structure without a disk). Due to the disk shadow the temperatures near the midplane of the structure are below $T_\mathrm{sub}(\mathrm{CO})$ and, similar to the disk component, CO freezes out quickly. However, the vertical density gradient in this region is much flatter compared to the disk and regions which are not in the shadow of the disk contribute equally to $\epsilon_\mathrm{avg}(\mathrm{CO})$. Nevertheless, the disk causes some depletion of CO within the radial CO ice line of the envelope. A 1D envelope model would not show such an additional depletion and the average CO abundance would reach the canonical value of $\epsilon_\mathrm{avg}(\mathrm{CO})\approx2\times10^{-4}$ within the envelope CO ice line. 

Beyond $r\gtrsim 300\,\mathrm{au}$ the dust temperature is below $T_\mathrm{sub}(\mathrm{CO})$ and CO can freeze out throughout the whole structure. Due to the lower densities in this zone (e.g. $n_\mathrm{<H>}(r=300\,\mathrm{au})\approx5\times10^6\,\mathrm{cm^{-3}}$ and $n_\mathrm{<H>}(r=2500\,\mathrm{au})\approx10^5\,\mathrm{cm^{-3}}$) the \mbox{freeze-out} timescale increases significantly. The delayed freeze-out is nicely seen in the averaged abundance profiles and in the 2D contour plots. Using Eq.~(\ref{eq:freezetimescale}) the \mbox{freeze-out} timescales at $\approx300\,\mathrm{au}$ and $\approx2500\,\mathrm{au}$ are $\approx400\,\mathrm{yr}$ and $\approx20\:000\,\mathrm{yr}$ respectively. The difference in the timescale for these two points is mainly a result of the density gradient; the temperature varies only by a factor of two, the density by a factor of $\approx50$. As as consequence we see an inside-out freeze-out of CO similar to the 1D models of \citet{Visser2015c} (see also Appendix~\ref{sec:visser}).

As Fig.~\ref{fig:CI_COabun} shows, it is not trivial to provide a single number for a radial CO ice line in a complex 2D structure. The picture is further complicated by the slow evolution of the CO abundance in the post-burst phase. However, it will be beneficial for the rest of the paper to define two distinct locations for the radial CO ice lines. The first is the CO ice line in the quiescent phase which is at $R_\mathrm{Q}(\mathrm{CO\#})\approx300\,\mathrm{au}$ (\# stands for ice), the second is the location of the ice line during the burst at $R_\mathrm{B}(\mathrm{CO\#})\approx2500\,\mathrm{au}$. These two radial CO ice lines roughly correspond to the location of the $T_\mathrm{dust}=27\,\mathrm{K}$ (the CO sublimation temperature) contours seen in Fig.~\ref{fig:structure} but are also clearly visible in Fig.~\ref{fig:CI_COabun}. We do not consider the CO ice line in the disk for the further discussion, because the main action in the post-burst phase, concerning the CO abundance, is happening in regions $R_\mathrm{Q}(\mathrm{CO\#})\lesssim r\lesssim R_\mathrm{B}(\mathrm{CO\#})$.

Comparing our model to the spherical symmetric 1D model of \citet{Visser2015c} (see Appendix~\ref{sec:visser}) shows that the evolution of the CO abundance with time is qualitatively speaking similar in both models. Although adding a disk component has a significant impact on the temperature structure, the density gradient on large scales (the envelope) is not affected. As the freeze-out timescale is mainly determined by the density, the time-evolution of the CO gas phase abundance in the outer regions of the envelope is therefore not strongly affected by the presence of a disk (see also Sect.~\ref{sec:impactstruc}). However, as already discussed, the disk has an impact on the actual freeze-out timescale in the inner region of the structure and on the detailed location of the CO ice line(s) in the envelope structure which are  relevant for the quantitative interpretation of observations. 
\begin{figure*}
\centering
  \includegraphics{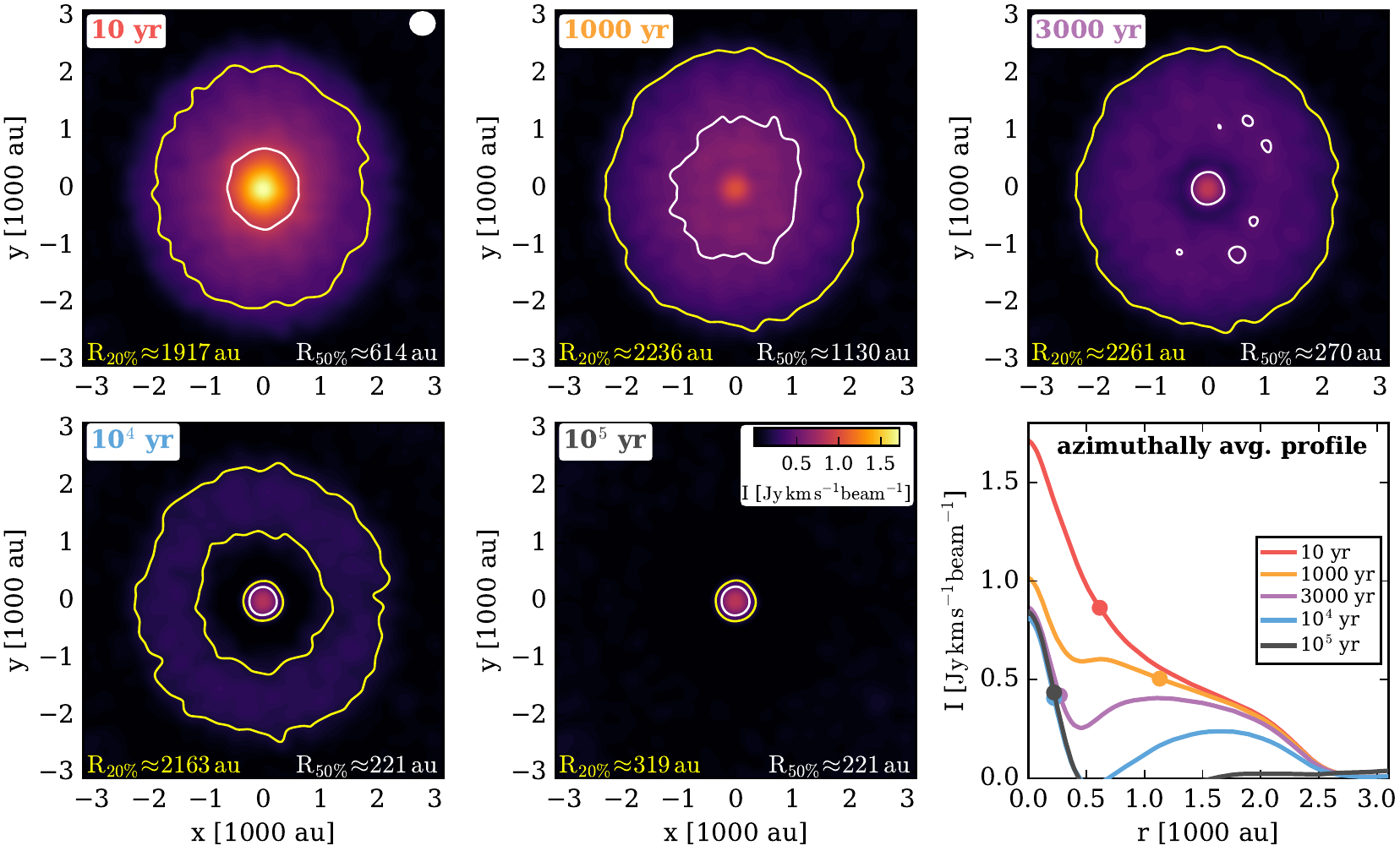}  
  \caption{\mbox{$\mathrm{C^{18}O}\:J\!=\!2\!-\!1$} ALMA simulations for the representative Class~I model. The first five panels from top left to the bottom right show integrated intensity maps for the post-burst phase at 10, 1000, 3000, 10$^4$ and 10$^5\,\mathrm{yr}$ years after the end of the burst (see also Fig.~\ref{fig:CI_COabun}). The target is seen face-on (looking down the outflow cavity, $\mathrm{inclination}=0\degr$). The white ellipse in the top left plot shows the synthetic beam ($1.81\mathrm{\arcsec}\times1.65\mathrm{\arcsec}$). The linear scale for the intensity is shown in the bottom centre plot. The white and yellow contour lines show 50\% and 20\% of the peak intensity level for each map. The plot at the bottom right shows the azimuthally averaged intensity profiles for the different times. The dots in this plot mark the full width half maximum of the profile ($R_{50\%}$).}
  \label{fig:almasim0}
\end{figure*}
\subsection{ALMA simulations}
\label{sec:almasim}
To study the impact of the chemical evolution in the \mbox{post-burst} phase on observables we present synthetic observations for the \mbox{$\mathrm{C^{18}O}\:J\!=\!2\!-\!1$} spectral line using proper line radiative transfer (Sect.~\ref{sec:syntheticobs}) and CASA/ALMA simulations (see Appendix~\ref{sec:alma_method} for details). We use \mbox{$\mathrm{C^{18}O}\:J\!=\!2\!-\!1$} for two main reasons. Firstly, CO has a low sublimation temperature, therefore CO sublimates also in the outer regions of the structure where the timescale for freeze-out is the longest. This increases the probability to detect extended CO emission long after the burst \citep{Jorgensen2015b}. Secondly, choosing \mbox{$\mathrm{C^{18}O}\:J\!=\!2\!-\!1$} allows for comparison of our results to the 1D models of \citet{Jorgensen2015b} as they used the measured extent of the \mbox{$\mathrm{C^{18}O}\:J\!=\!2\!-\!1$} emission to identify post-burst objects. 

Figure~\ref{fig:almasim0} shows \mbox{$\mathrm{C^{18}O}\:J\!=\!2\!-\!1$} intensity maps for the same times as shown in Fig.~\ref{fig:CI_COabun}. The target is seen at an inclination of $0\degr$ (i.e. face-on; the observer looks down the outflow cavity along the z axis). The last panel in Fig.~\ref{fig:almasim0} shows the azimuthally averaged radial intensity profiles. To indicate the extent of the emission we show contours for $50\%$ and $20\%$ of the peak intensity. The radius for the $50\%$ contour is also marked in the averaged intensity profiles. The radius of the $50\%$ contour, $R_\mathrm{50\%}$, can be seen as a measure for the extent of the CO emission. We follow this approach as in observational studies often the full width at half maximum (FWHM) of a Gaussian fitted to the observation, is used to measure the extent of emission (e.g. the radius is given by FWHM/2, see also Sect.~\ref{sec:fitdiscussion}). However, the $R_\mathrm{50\%}$ radius shown in Fig.~\ref{fig:almasim0} is not necessarily equal to the FWHM/2 of a fitted Gaussian as we use here the full profile. 

From the azimuthally averaged radial intensity profiles in Fig.~\ref{fig:almasim0} one can see that the observations nicely trace the evolution of the gas-phase CO as discussed in Sect.~\ref{sec:COabundance}. Due to the faster freeze-out of CO in the inner regions a dark gap appears in the intensity maps. This gap is already visible at $t=1000\,\mathrm{yr}$ and grows with time until it disappears at $t\gtrsim10\:000\,\mathrm{yr}$, when all the CO released into the gas phase during the burst, is frozen-out again.

Another interesting aspect is the evolution of the peak intensity. As seen in the panel for the radial profiles, the peak intensity reaches its final or quiescent level already at $t\approx1000\,\mathrm{yr}$. We note that for this particular simulation the disk is not resolved (the beam size corresponds to $300-350~\mathrm{au}$ at a distance of $200\,\mathrm{pc}$). Nevertheless, the peak intensity is mostly determined by emission from and close to the disk if the structure is seen face-on. Therefore the peak intensity evolves on a timescale of $100$ to $1000~\mathrm{yr}$. As a consequence of the differential freeze-out also the apparent extent of the emission is affected. As nicely seen in the averaged profiles, the $R_\mathrm{50\%}$ radius at $t=1000\,\mathrm{yr}$ is larger than at $t=10\,\mathrm{yr}$.
\begin{figure*}
\centering
  \includegraphics{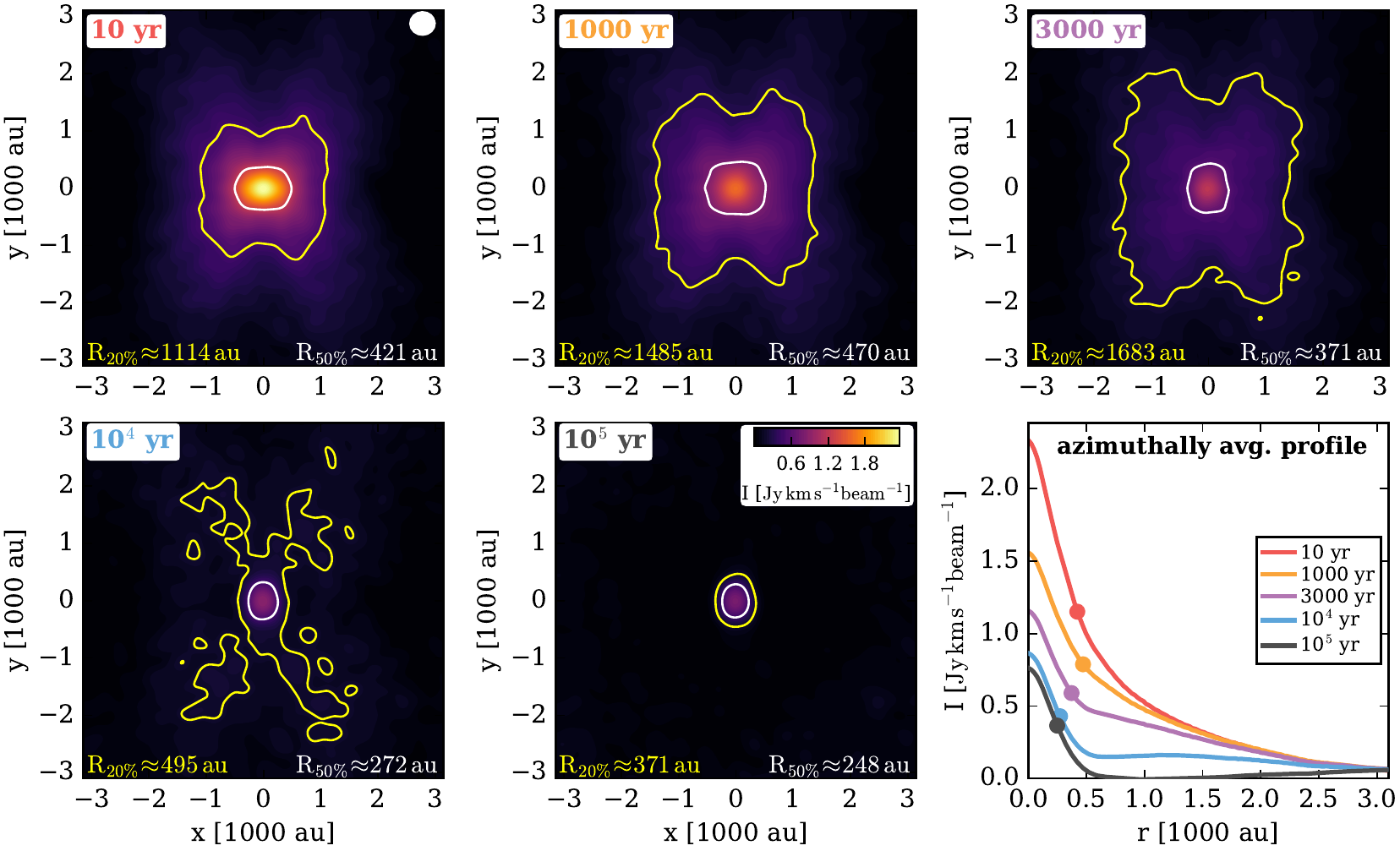}  
  \caption{The same as Fig.~\ref{fig:almasim0} but for an inclination of 90\degr
  (edge on, perpendicular to the outflow axis).}
\label{fig:almasim90}
\end{figure*}
\begin{figure*}
\centering
  \includegraphics{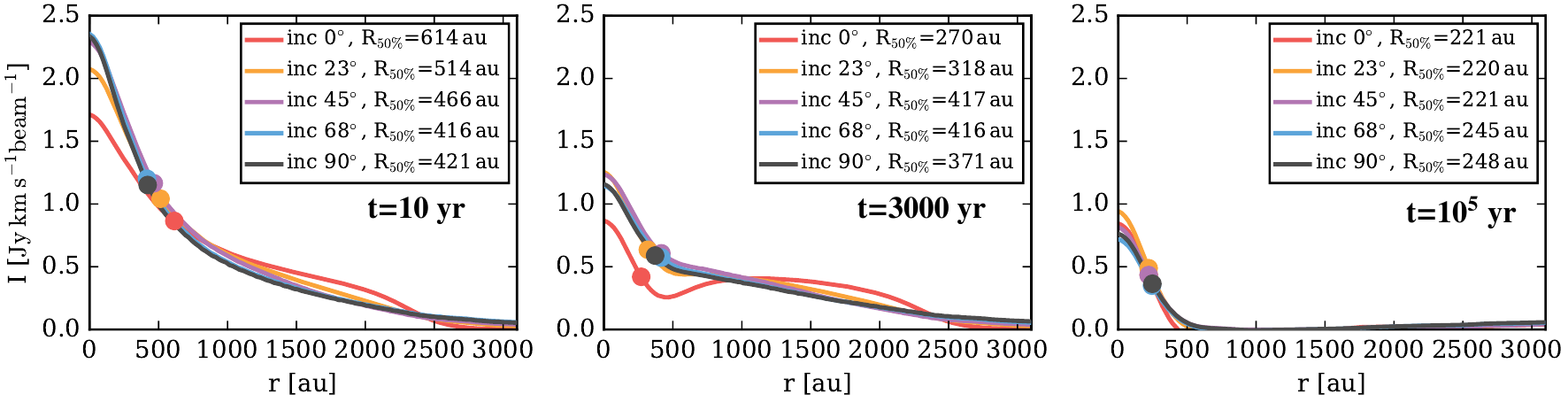}
  \caption{Impact of inclination on the radial intensity profiles. Shown are azimuthally averaged radial \mbox{$\mathrm{C^{18}O}\:J\!=\!2\!-\!1$} intensity profiles for models with different inclinations (coloured lines in each panel) at three different times after the burst (different panels). The large dot in each profile indicates the $R_\mathrm{50\%}$ radius corresponding to the value given in the legend.}
  \label{fig:inc_rp}
\end{figure*}
In Fig.~\ref{fig:almasim90} we show the same model as in Fig.~\ref{fig:almasim0} but the target is now seen at an inclination of 90\degr (edge-on, perpendicular to the outflow axis). The main difference to the face-on view is the absence of a gap and the X-shape of the emission in the post-burst phase, best seen in the panel for $t=10^4\,\mathrm{yr}$. The reason for the absence of the gap is that for inclined targets one mainly sees the CO on large scales which dominates the emission along the line of sight. Therefore a detection of the gap is only likely for targets seen nearly face-on, where one can peek down the outflow cavity. In our models the gap is only visible for inclinations $\lesssim23\degr$ (see Fig.~\ref{fig:almasiminc}). 

The X-shape of the emission is a consequence of the outflow cavity and again the different freeze-out timescales. In regions with the outflow cavity one sees simply less material (as the cavity is empty) and therefore also weaker emission. Perpendicular to the outflow cavity axis we see more material, but due to higher densities close to the midplane of the structure ($y=0$ in Fig.~\ref{fig:almasim90}) CO freezes out faster than close to the outflow walls. The higher densities close to the midplane are due to the rotationally flattened structure and the disk. Further in the disk shadow the temperature is cooler and the CO ice line close to the midplane is located a smaller radii compared to the regions close to the outflow walls (see Fig.~\ref{fig:CI_COabun}). As a consequence of these effects the \mbox{X-shape} of the emission is most pronounced at high inclinations (see also Fig.~\ref{fig:almasiminc}).

In Fig.~\ref{fig:inc_rp} we show azimuthally averaged intensity profiles at three different times (including the quiescent state) after the burst, where in each panel models with different inclinations are shown. In the quiescent state ($t=10^5\,\mathrm{yr}$) inclination has only a marginal impact on the resulting intensity profiles and the $R_\mathrm{50\%}$ radii vary only by about 10\%. For times shortly after the burst ($t=10\,\mathrm{yr}$ and $t=3000\,\mathrm{yr}$) the situation is more complex. For $t=10\,\mathrm{yr}$ the extent of the CO emission is larger for smaller inclinations whereas for $t=3000\,\mathrm{yr}$ the opposite is true. The reason for this is the freeze-out of CO in the inner regions which affects the peak intensity and therefore also the $R_\mathrm{50\%}$ radius but also optical depth effects play a role here (see Sect.~\ref{sec:opticaldepth}).
However, even for inclined targets the peak intensity evolves on shorter timescales than the extended emission which is a consequence of the different freeze-out timescales. Similar to the face-on models the measured $R_\mathrm{50\%}$ radius can be larger in the post-burst face than during or shortly after the burst.

\begin{figure*}
\centering
  \includegraphics{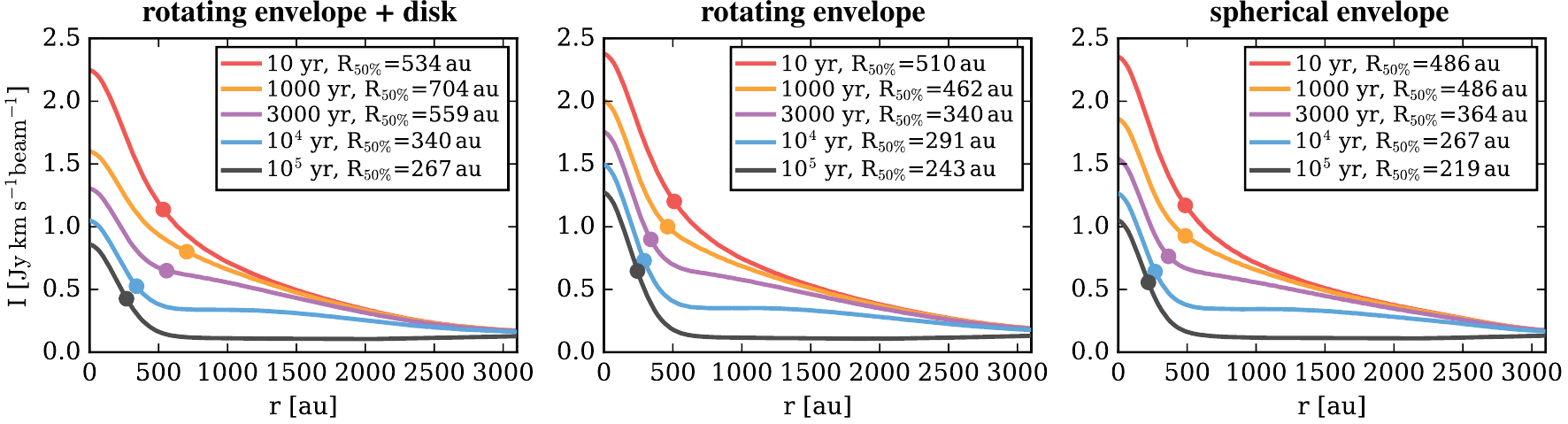}  
  \caption{Impact of structure on the radial intensity profiles. Shown are azimuthally averaged radial intensity profiles for \mbox{$\mathrm{C^{18}O}\:J\!=\!2\!-\!1$} for models with an inclination of 45$\degr$ at different times in the post-burst phase. Density structures from left to right: rotating envelope with a disk component, rotating envelope without a disk, a spherical symmetric envelope (all models include an outflow cavity). The large dot in each profile indicates the $R_\mathrm{50\%}$ radius corresponding to the value given in the legend.}
  \label{fig:struc_rp}
\end{figure*}
The presented ALMA simulations clearly show that the inside out-freeze out produces distinct observational signatures (such as the gap) in spatially resolved images of \mbox{$\mathrm{C^{18}O}\:J\!=\!2\!-\!1$}. The main requirement for real interferometric observations is that the different spatial scales are properly captured and that the large scale emission is not filtered out. Although we presented here only ALMA simulations also other existing (sub)mm interferometers like IRAM-PdBI/NOEMA (IRAM Plateau de Bure Interferometer/NOrthern Extended Millimeter Array) and SMA (Submillimeter Array) are capable of performing such observations (e.g. \citealt{Jorgensen2015b,Anderl2016}).
\subsection{Impact of structure}
\label{sec:impactstruc}
Additionally to our representative Class~I model we performed the same burst simulations for `simpler' structures, namely a rotating-envelope model without a disk and a spherical symmetric model. The main parameters, like the stellar properties, the outflow cavity and the extension of the models are the same. The main difference lies in the radial density profiles. For the rotating-envelope model the slope of the density gradient flattens towards the centre; in the spherical model the radial density distribution is simply proportional to $r^{-1.5}$ (i.e. setting the centrifugal radius $R_\mathrm{c}$ in Eq.~\ref{eqn:envstruc3} to zero).

In Fig.~\ref{fig:struc_rp} we show a comparison of the three different structure models. Shown are the averaged radial intensity profiles at several times after the burst. The inclination is $45\degr$ (see Fig.~\ref{fig:rp_struciall} for the other inclinations). To produce those synthetic observations we did not perform full ALMA simulations but present simple beam-convolved simulations, with the same beam size as we used for the ALMA simulations. The beam-convolved simulations represent the radial intensity profiles from the ALMA simulations very well (see Appendix~\ref{sec:alma_method}) and are mainly used to save computational resources.   

At first sight, the evolution of the radial intensity profiles look quite similar for all three structure models, but there are also distinct differences. In particular, the evolution of the peak intensity happens on different timescales. For the envelope+disk model the peak intensity drops quickly with time due to the fast freeze-out ($\approx 100\,\mathrm{yr}$) in the midplane of the disk. The two other structure models show a slower evolution of the peak intensity, where the rotating envelope model shows the slowest due to the flattening of the density profile towards the centre. 

In Fig.~\ref{fig:struc_rp} we also indicate the $R_\mathrm{50\%}$ radius and give the actual value in the legend of each panel. In the envelope+disk model $R_\mathrm{50\%}$ is larger at all times than in the two other structure models. The reason for this is actually the lower peak intensity and not the extension of the CO emission. Due to the higher density and lower temperatures in the disk midplane, CO freezes out quicker and the averaged CO abundance is lower compared to the structure models without a disk.

The comparison of the three structure models shows that on large scales the evolution of the radial intensity profiles are similar. However, due to the different density structures in the inner regions ($r<500\,\mathrm{au}$) the peak intensity evolves on different timescales. As a consequence, the actual measured extent of the emission is larger for structures with a steeper density gradient or a high density component, such as a disk, in the inner regions. Such effects are relevant for the quantitative interpretations of CO observations in the post-burst phase and can only be properly captured by 2D models like the one presented here.
\begin{figure*}
\centering
  \includegraphics{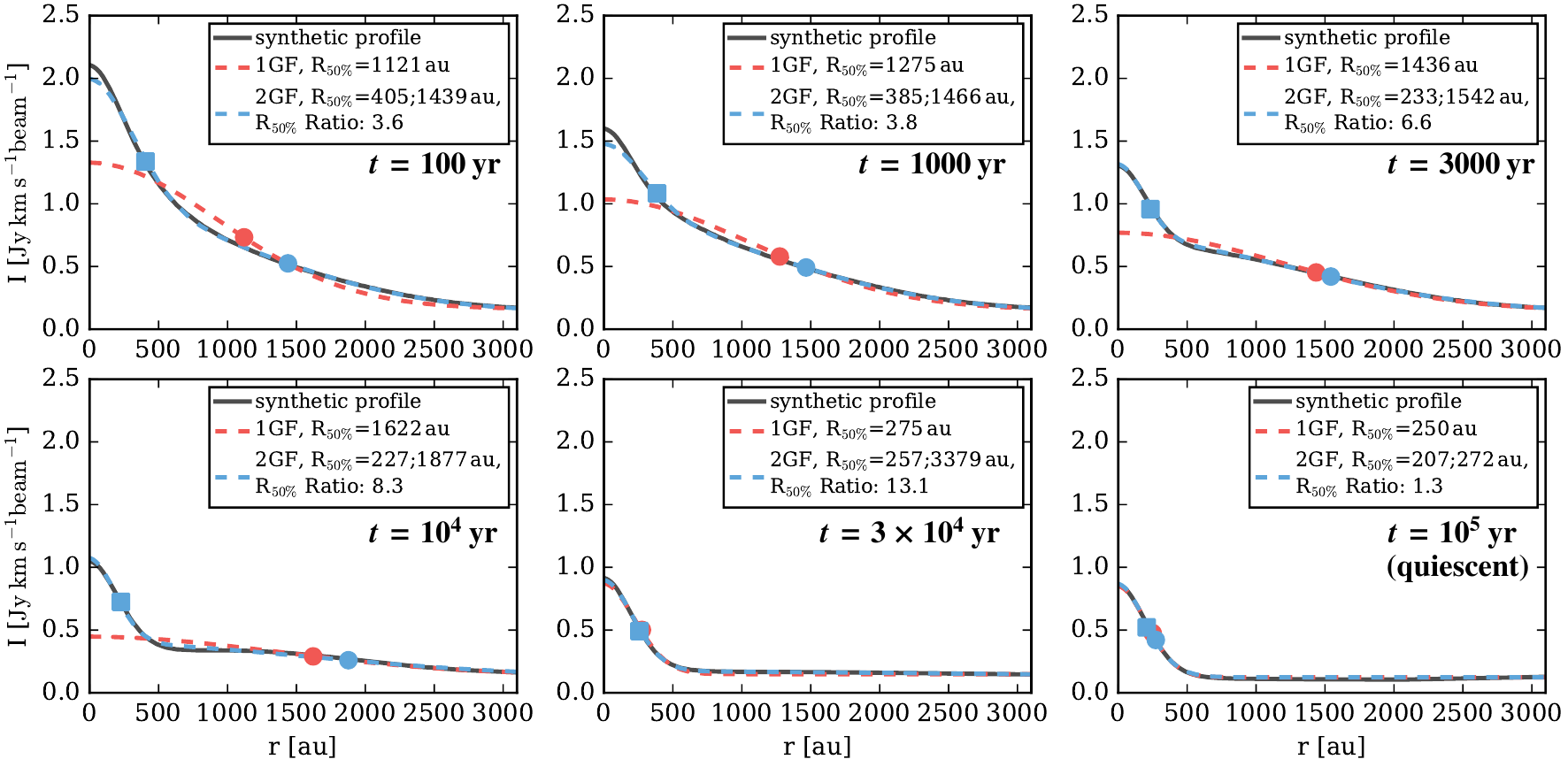}  
  \caption{\mbox{$\mathrm{C^{18}O}\:J\!=\!2\!-\!1$} radial intensity profiles derived from fitting the synthetic observations. Shown are the results for an inclination of $45\degr$ at five different times (given in the panels) in the post-burst phase and in the quiescent phase (last panel). Each individual panel shows the `real' beam convolved simulations (black solid line), a fit using one Gaussian (1GF, red dashed line) and a fit using two Gaussians (2GF, blue dashed line). In the legend the FWHM/2 ($R_\mathrm{50\%}$) value for the fitted Gaussian(s) is denoted (indicated by the big dots). For the 2GF both radii (for each Gaussian) and the ratio of the two radii are provided. For comparison, the actual CO ice lines in the quiescent phase and in the burst phase are located at $R_\mathrm{Q}(\mathrm{CO\#})\approx300\,\mathrm{au}$ and $R_\mathrm{B}(\mathrm{CO\#})\approx2500\,\mathrm{au}$, respectively (see Sect.~\ref{sec:COabundance} and Fig.~\ref{fig:CI_COabun}).}
  \label{fig:almafit}
\end{figure*}
\section{Discussion}
\label{sec:discussion}
\subsection{A model independent method to identify post-burst sources}
\label{sec:fitdiscussion}
To measure the radial extent of emission it is common to use the FWHM of a 1D/2D Gaussian fitted to the observational data. Such an approach is also used in \citet{Jorgensen2015b} to estimate the extent of \mbox{$\mathrm{C^{18}O}\:J\!=\!2\!-\!1$} for a sample of Class~0/I sources to identify targets that experienced a recent burst. \citet{Jorgensen2015b} find extended \mbox{$\mathrm{C^{18}O}\:J\!=\!2\!-\!1$} emission, with respect to the extent expected from the currently observed source luminosity, for half of their targets.

It is interesting to see what radial extent for CO would be measured from our models,  using a similar fitting procedure. We fit either one or two Gaussians to the beam convolved images using the CASA task \texttt{imfit}. The fit also includes a zero level offset to account for the background $\mathrm{C^{18}O}$ emission on large scales. To compare the fitting results to the synthetic observations we use again azimuthally averaged radial intensity profiles, produced in the same way as for the synthetic observations.

In Fig.~\ref{fig:almafit} we show the derived radial intensity profiles at six different times after the burst. Each panel of Fig.~\ref{fig:almafit} shows the synthetic profile and the profiles derived from the two different fitting methods (1GF and 2GF, respectively). Further the measured $R_\mathrm{50\%}$ radii are given (correspond to FWHM/2 of the fitted Gaussians). In the case of the two-component fit (2GF) both measured radii for the individual Gaussians and the ratio of the two radii are denoted.

Comparing the $R_\mathrm{50\%}$ radius in the quiescent phase to the measured radii in the post-burst phase shows that the single Gaussian (1GF) indeed is a reliable method to identify extended CO emission in the post-burst phase. However, it is interesting to see that $R_\mathrm{50\%}$ increases with time until all CO sublimated during the burst is frozen-out again ($t\approx3\times10^4\,\mathrm{yr}$). The reason for this is that the fit is more sensitive to the extended emission (larger area) and that the the peak of the profiles evolves on a shorter timescale than the most extended emission (see Sections \ref{sec:almasim} and \ref{sec:impactstruc}). It is also apparent that the emission at small radii is not well fitted with the 1GF in the post-burst phase, only in the quiescent phase the emission is reasonably well represented by a single Gaussian.

The two Gaussian fitting procedure (2GF) fits the synthetic profiles in the post-burst phase significantly better than the single Gaussian fits, now also the emission on small scales is fitted well. Typically the $\chi^2$ value of the 2GF is a factor of approximately four lower than for the 1GF for $t<3\times10^4\,\mathrm{yr}$. $R_\mathrm{50\%}$ for the extended emission (the larger of the two radii indicated in Fig.~\ref{fig:almafit}) derived from the two-component fit is usually slightly larger compared to the 1GF, but considering uncertainties those radii are quite similar. In the case of very weak extended emission or in the quiescent phase, the 2GF either fails (see Appendix~\ref{sec:fittingexamples}) or the quality of the 2GF and 1GF is nearly identical.

It is not surprising that fitting two components provides better results than a single component fit, simply because the 2GF has more free parameters. Besides this mathematical argument there are also physical reasons why a two component fit is a good representation of the post-burst emission pattern. The CO emission in the post-burst phase can be separated into two components. One component corresponds to the emission coming from within the radial CO line corresponding to the current (quiescent) temperature structure ($R_\mathrm{Q}(\mathrm{CO\#})\approx300\,\mathrm{au}$, see Sect.~\ref{sec:COabundance}). This component exists in all phases: burst, post-burst and quiescent phase. The second component corresponds to the extended emission coming from the region \mbox{$R_\mathrm{Q}(\mathrm{CO\#})\lesssim r \lesssim R_\mathrm{B}(\mathrm{CO\#})$}. In the post-burst phase CO freezes out between these two ice lines where the longest freeze-out timescale is close to $R_\mathrm{B}(\mathrm{CO\#})$. In the actual observation we see (depending on the viewing angle) a superposition of these two components, it is therefore advantageous to actually use also two components to fit such observations.

Besides the better quality of 2GF compared to the 1GF, the 2GF fitting procedure has several further advantages:
\begin{itemize}
  \item with the 2GF procedure one obtains information about the CO ice line in the quiescent and burst-phase as two radii are measured. The measured smaller radius corresponds to the quiescent CO ice line and the larger radius to the burst CO ice line. Of course both quantities are only a rough estimate and especially the measured quiescent CO ice line should be seen as an upper limit (i.e. depending on the spatial resolution available)
  \item the ratio of the two radii provides some rough indication for the time since the last burst. The ratio usually increases with time as can be seen in Fig.~\ref{fig:almafit}. The ratio increases because of the different freeze-out timescales in the inner region and outer region. Emission close to $R_\mathrm{B}(\mathrm{CO\#)}$ is seen for longer than the emission close to $R_\mathrm{Q}(\mathrm{CO\#})$.
  \item for the quiescent phase the 1GF is actually a better representation of the observations as only one component is seen. For the quiescent phase the 2GF either fails at all or performs equally well as the 1GF fit. Only for the post-burst profile the 2GF is superior (typically at least a factor four lower $\chi^2$ values). By using the two methods to fit real observations it is therefore possible to identify post-burst targets in a model independent way. For the 1GF the measured extent needs to be compared to a model predicting the actual extent for the current measured source luminosity. The 1GF approach depends in particular on the binding energy of CO which defines the location of the CO ice line(s) (see \citet{Jorgensen2015b} for a discussion). This is not the case for the 2GF approach. In that case it is possible to identify post-burst sources by simply comparing the quality of the 2GF to the 1GF. A significantly better quality of the 2GF already indicates that the object is currently in the post-burst phase. Therefore the 2GF method does not require any detailed modelling to identify post-burst objects. 
\end{itemize}
We also tested the 2GF method for the different structure models, different inclinations and different spatial resolutions (see Appendix~\ref{sec:fittingexamples} for further examples). Although the absolute numbers for the measured radii can vary, the main arguments in favour of the 2GF are also valid for those models. However, it still must be shown how well the procedure works with other models (in particular chemical models) and subsequently with real observational data. However, the main physical argument for the two Gaussian fitting procedure is actually the inside-out freeze-out of CO in the zone between the quiescent and burst CO ice lines. This is a very robust chemical result as it is mainly based on the adsorption timescales and is also seen in other models \citep[e.g.][]{Visser2015c,Vorobyov2013f}. The 2GF method is therefore a robust and consistent way to identify post-burst targets and to derive statistically relevant information such as the strength and frequencies of bursts.
\subsection{CO extent versus bolometric luminosity}
\label{sec:lumrad}
\begin{figure}
\centering
  \resizebox{\hsize}{!}{\includegraphics{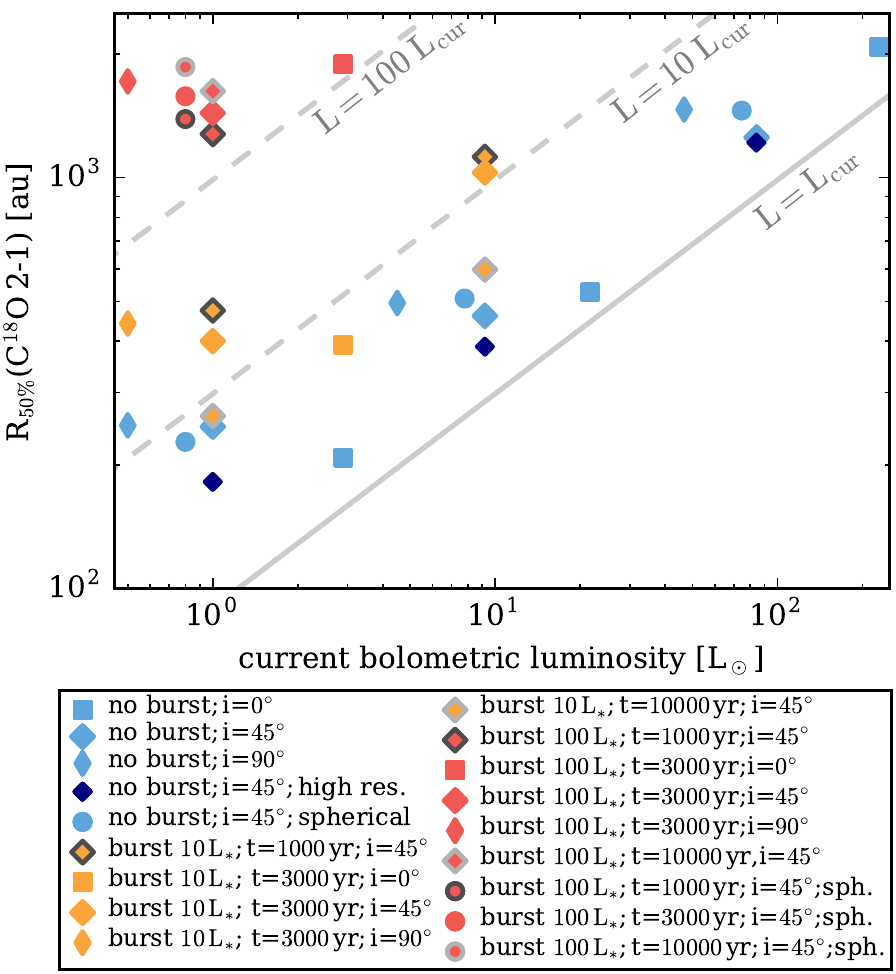}}
  \caption{Measured \mbox{$\mathrm{C^{18}O}\:J\!=\!2\!-\!1$} extent versus current bolometric luminosity. The blue symbols show models in quiescent state (no burst) with protostellar luminosities $L_\mathrm{*}$ of 1, 10 and 100$\,\mathrm{L_\sun}$. The red symbols are for strong burst models with burst luminosities of $L_\mathrm{*,B}=100\times L_\mathrm{*}$. The orange symbols show weak burst models with $L_\mathrm{*,B}=10\times L_\mathrm{*}$. For both groups of burst models the results in the post-burst phase are shown. The edges of the symbols indicate the time past since the end of the burst of $1000\,\mathrm{yr}$ (black edge), $3000\,\mathrm{yr}$ (no edge) and $10\,000\,\mathrm{yr}$ (grey edge). The different shapes of the symbols indicate different inclinations of $0\degr$ (square), $45\degr$ (diamond) and $90\degr$ (thin diamond). For all models a beam size of $1.81\arcsec\times1.65\arcsec$ was used, except for the models indicated by the dark blue symbols where a beam size of $0.5\arcsec\times0.5\arcsec$ (high res.) was used. The grey solid and dashed lines are the model results from \citet{Jorgensen2015b} for a 1D spherical model of a Class~0 protostellar envelope with varying protostellar luminosities.}
  \label{fig:lumrad}
\end{figure}

As mentioned above, \citet{Jorgensen2015b} identified post-burst candidates by relating the measured extent of \mbox{$\mathrm{C^{18}O}\:J\!=\!2\!-\!1$} emission to the currently measured bolometric luminosity of the target. To compare our results to this approach we show in Fig.~\ref{fig:lumrad} a similar CO extent versus current bolometric luminosity plot as in \citet{Jorgensen2015b} and populate this plot with the values derived from our models.

We want to emphasize that the models of \citet{Jorgensen2015b} are very different to the models presented in this paper. They use 1D models with a spherical power-law density distribution $\rho\propto r^{-1.5}$ and calculate the temperature structure with a radiative transfer code for a range of luminosities. The CO abundance is modelled with a simple step function where the CO abundance is decreased by two orders of magnitude for $T<30\,\mathrm{K}$. To measure the extent of CO in those models they convolved the synthetic images with a $\approx2\arcsec$ beam and fitted a single 1D Gaussian in the visibility plane. The FWHM/2 of this fitted Gaussian gives the radius of the CO emission (the $R_\mathrm{50\%}\mathrm{(C^{18}O\:2\!-\!1)}$ axis in Fig.~\ref{fig:lumrad}). Because of all these differences we do not aim for a direct comparison of the models but rather to use the results of \citet{Jorgensen2015b} as a reference frame.

To populate Fig.~\ref{fig:lumrad} we calculate the bolometric luminosity of the model by integrating the synthetic spectral energy distribution and measure the CO extent by fitting our synthetic \mbox{$\mathrm{C^{18}O}\:J\!=\!2\!-\!1$} images as already described in Sect.~\ref{sec:fitdiscussion}. To be more consistent with \citet{Jorgensen2015b} we use here the $R_\mathrm{50\%}\mathrm{(C^{18}O\:2\!-\!1})$ derived from the single Gaussian fitting procedure. We want to note that the measured bolometric luminosity actually depends on inclination and does not necessarely represent the true source luminosity. If a Class~I object is seen face-on one sees the maximum protostellar flux and scattered emission where for the edge-on case the protostar is usually obscured (see \citealt{Whitney2003} for a detailed discussion). For our models presented here the measured bolometric lumonisity is typically about factor of two to three higher for the face-on case and a factor of two lower for the edge-on case, compared to the true source luminosity. These values are in good agreement with the results of \citealt{Whitney2003}. 

In Fig.~\ref{fig:lumrad} we show various different models. The first group of models are models without a burst (quiescent). We show models with a protostellar luminosity $L_\mathrm{*}$ of $1$, $10$ and $100\,\mathrm{L_{\sun}}$. For each of those models we plot the measured CO extent versus the actual bolometric luminosity (calculated from the synthetic SEDs) for three different inclinations. Further we show no-burst models using only a spherical structure (no disk component) and models where the CO extent was derived from synthetic observations with a higher spatial resolution (beam with $0.5\arcsec\times0.5\arcsec$; corresponds to $\approx100\,\mathrm{au}$ resolution at a distance of $200\,\mathrm{au}$).

The second group of models corresponds to strong burst models with a burst luminosity 100 times the quiescent luminosity ($L_\mathrm{*,B}=100\times L_\mathrm{*}$). These models are the same as discussed in Sect.~\ref{sec:results}. For those models we show the measured CO extent in the post-burst phase at three different times, for different inclinations and also for the spherical structure model. We note that for the post-burst models the current measured bolometric luminosity is equal to the quiescent phase as the burst is over and the temperature structure already reached its quiescent state. The third group of models represents weak bursts, where the quiescent stellar luminosity is increased only by a factor of ten during the burst ($L_\mathrm{*,B}=10\times L_\mathrm{*}$).

A close inspection of Fig.~\ref{fig:lumrad} unveils several interesting aspects of our 2D radiation thermo-chemical model 
\begin{itemize}
\item \emph{inclination:} A glance at the models without a burst shows that the impact of inclination is twofold. The different inclinations produce a scatter along the luminosity axis although the physical structure and properties of the models are otherwise the same (see discussion above). The different inclinations cause also some scatter for the measured CO extent, but compared to the bolometric luminosity this is rather limited. We note that those effects are a natural outcome of our 2D model and cannot be properly captured with 1D models. However, due to our simplistic model for the outflow cavity (i.e. it is empty and we consider only one opening angle for the cavity) our here presented models can only provide a rough estimate for the impact of inclination.
\item\emph{structure:} Comparing the spherical models to the disk+envelope models shows again that the details of the structure are not particularly significant on large scales (see also Sect.~\ref{sec:impactstruc}). 
\item\emph{spatial resolution:} High spatial resolution observations ($0.5\arcsec\times0.5\arcsec$ beam) provide obviously more accurate results for objects with a small CO extent, whereas for objects with large CO extent the measured radii are nearly identical to the radii of the low resolution models ($1.81\arcsec\times1.65\arcsec$ beam).
\item\emph{strong bursts ($100\times L_\mathrm{*}$)}: As already discussed in previous sections, the extent of the CO emission appears larger at later times after the burst. Fig.~\ref{fig:lumrad} clearly shows that strong recent bursts are easily detectable for $\mathrm{\approx10\:000\,\mathrm{yr}}$ after the end of the burst. Although the measured radii for the CO extent vary slightly with time the radii are at least a factor of approximately four larger compared to the CO extent expected from the current bolometric luminosity. We want to emphasize that for such strong bursts it is important to not filter out large scale structures in interferometric observations, otherwise such post-burst targets would not be detected (see also \citealt{Jorgensen2015b}).
\item\emph{weak bursts ($10\times L_\mathrm{*}$)}: For the weak burst models with a quiescent stellar luminosity of $L_\mathrm{*}=10\,\mathrm{L_{\sun}}$ and and a burst luminosity of $L_\mathrm{*,B}=100\,\mathrm{L_{\sun}}$ the measured CO extent is slightly smaller than for the strong burst models which also have $L_\mathrm{*,B}=100\,\mathrm{L_{\sun}}$. The reason for this is that the contrast between the quiescent component in the inner region, which is more extended for $L_\mathrm{*}=10\,\mathrm{L_{\sun}}$, and the extended post-burst component of the CO emission is weaker compared to strong burst models. In particular the peak luminosity is higher resulting in a slightly narrower width of the fitted Gaussian. In contrast to the strong burst models the weak burst models do not indicate extended CO emission at $t=10\,000\,\mathrm{yr}$ after the burst (i.e. $R_\mathrm{50\%}$ is similar to the quiescent state). One reason for this is, again, the weaker contrast between the quiescent and extended emission components the other is the freeze-out timescale. In the weak burst model with $L_\mathrm{*}=1\,\mathrm{L_{\sun}}$ and $L_\mathrm{*,B}=10\,\mathrm{L_{\sun}}$ the CO ice line during the burst is at smaller radii ($R_\mathrm{B}\mathrm{(CO\#)}\approx800\,\mathrm{au}$) where the freeze-out timescale is about a factor of three shorter than in the corresponding strong burst models ($R_\mathrm{B}\mathrm{(CO\#)}\approx2500\,\mathrm{au}$). Generally speaking, weak bursts are harder to detect in the post-burst phase and therefore the detection probability for weak bursts decreases significantly. Such limitations need to be considered for deriving statistical quantities such as burst frequencies from post-burst observations. 
\end{itemize}
After having discussed the details of Fig.~\ref{fig:lumrad}, a more global view of Fig.~\ref{fig:lumrad} shows that our model results are qualitatively in good agreement with the results of \citet{Jorgensen2015b}. Although there are quantitative differences in the models, which are not surprising as we use a different structure and chemical model, the general agreement is certainly a strong argument in favour of the CO extent method. The main advantage of using CO as a \mbox{post-burst} tracer are the expected long-freeze out timescale which allow to detect bursts up to $10\,000\mathrm{s}$ of years after the actual end of the burst \citep{Visser2015c,Jorgensen2015b}.
\subsection{Further considerations}
\subsubsection{Dynamical evolution and outflows}
\label{sec:dynamicalevol}
We assume here a steady-state structure and consequently ignore any dynamical evolution of the system. The issue here is that the CO abundance is out of equilibrium with the temperature structure. 

For our representative Class~I model the free-fall timescale at $r=2500\,\mathrm{au}$ is $t_\mathrm{ff}\approx{20\:000\,\mathrm{yr}}$, which is actually comparable to the freeze-out timescale of $t_\mathrm{ads}\approx23\:000\,\mathrm{yr}$ at this distance. Close to the quiescent CO ice line at $r=300\,\mathrm{au}$ the timescales are $t_\mathrm{ff}\approx{800\,\mathrm{yr}}$ and $t_\mathrm{ads}\approx400\,\mathrm{yr}$ (now $t_\mathrm{ads}<t_\mathrm{ff}$). Considering those timescales, CO initially in the gas phase at $r=2500\,\mathrm{au}$ will have been frozen-out when it reaches $r=300\,\mathrm{au}$ where it will sublimate again (see also \citealt{Visser2009k}).

Nevertheless, as $t_\mathrm{ff}\approx t_\mathrm{ads}$, the dynamical evolution likely has an impact on our results. A parcel of gas located at $r=2500\,\mathrm{au}$ moving with the free-fall velocity would move inwards by $\approx130\,\mathrm{au}$ in $\approx1000\,\mathrm{yr}$. In the post-burst scenario this means that the burst CO ice line moves inward even if there would be no freeze-out at all (we assume here that CO outside of the burst CO ice line is mostly in the ice phase). This simple example should be seen as a worst case scenario, as we have ignored any rotational motion. Rotation will slow down the inward motion and the impact of dynamical infall on the CO ice line location would be less severe. However, our results for the expected measured CO extent in the post-burst phase should be seen as upper limits. On smaller scales the impact of dynamical evolution is less severe as there usually $t_\mathrm{ads}<t_\mathrm{ff}$ (see above). Although the dynamical evolution might reduce the timescale on which post-burst targets can be detected it does not affect our main conclusions.

This is also indicated by the hydrodynamic models of \citet{Vorobyov2013f}. They use the thin-disk approximation (averaged vertical quantities) to model the evolution of a protostellar system starting from the collapse up to the T~Tauri phase (see also \citealt{Vorobyov2010}). They model the dynamical evolution of CO, including adsorption and desorption processes, during and after accretion bursts. From their Fig.~3 one can see that their model shows similar features as presented here. In particular the radial gradient in the gas-phase CO abundance, resulting in a ring-like structure in our synthetic observations, can also be seen in their models. This provides further confidence that our results are not significantly affected by the dynamical evolution. 

Nevertheless, the dynamical timescale can vary from object to object (e.g different central masses, rotation of the envelope) and further investigations concerning the impact on observables in the post-burst phase are desirable (e.g. by producing synthetic observations from models like presented in \citealt{Vorobyov2013f}).

Although our model includes an outflow cavity, the outflow itself is not modelled at all. This is not necessarily an issue as long as $\mathrm{C^{18}O}$ emission from the envelope and disk is not polluted by emission from the outflow. Indeed observations indicate that $\mathrm{C^{18}O}$ traces mainly the envelope of embedded sources, in contrast to the more optically thick isotopologues $\mathrm{^{12}CO}$ and $\mathrm{^{13}CO}$, which commonly show high velocity wings in their spectral line profiles \citep[e.g][]{Frank2014,Dionatos2010}. Recent observations indicate that this is also the case for burst sources \citep[e.g.][]{Kospal2017,Ruiz-Rodriguez2017,Zurlo2017}.

In case outflows contribute to $\mathrm{C^{18}O}$ emission it should be possible to  disentangle the outflow and envelope emission components, as outflow velocities are higher than typical infall and rotation velocities of envelopes. Of course this is only possible for spectrally and spatially resolved observations and if the $\mathrm{C^{18}O}$ emission (in particular from the outflow) is mostly optically thin.

Outflows most likely also have an impact on the shape of the surrounding envelope structure, in particular in burst sources where strong outflows might be common \citep{Ruiz-Rodriguez2017,Zurlo2017}. Our here presented structure represents therefore a rather idealistic case as dynamical processes likely produce inhomogeneities in the density distribution. We want to emphasize here that our model primarily shows the impact of chemistry on the observables and that real observations will be to some extent also affected by dynamical processes.

\subsubsection{Optical depth effects}
\label{sec:opticaldepth}
It is commonly assumed that the \mbox{$\mathrm{C^{18}O}\:J\!=\!2\!-\!1$} line emission is on average optically thin in embedded sources \citep[e.g.][]{Jorgensen2015b,Anderl2016}. This is also the case for most regions in our model, at least in the quiescent phase. However, even in the quiescent phase in parts of the region around the disk \mbox{$\mathrm{C^{18}O}\:J\!=\!2\!-\!1$} becomes optically thick (at least in the line centre). This means that the innermost region, in particular the disk midplane, are to some extent obscured in the synthetic observations.

During the burst and shortly after the burst the optically thick region is much larger (up to $r\approx 1000\,\mathrm{au}$ depending on the viewing angle) due to the additional gas-phase CO in the outer regions of the structure. However, with time CO freezes out again and one can see deeper into the structure. This is also apparent from the synthetic images shown in Figs.~\ref{fig:almasim0} and \ref{fig:almasim90}. In the face-on images, the gap (close to the quiescent CO ice line) is seen as CO freezes out faster than at larger radii, consequently also the emission sooner becomes optically thin than at larger radii. If the object is inclined one sees more gas-phase CO along the line of sight from the outer region and therefore the inner regions are not seen as clearly. Although in parts of the structure the \mbox{$\mathrm{C^{18}O}\:J\!=\!2\!-\!1$} line emission is optically thick during and shortly after the burst the evolution of the CO freeze-out is still visible as the emission becomes optically thin with time. A comparison of the synthetic images and radial intensity profiles to the CO gas phase evolution shown in Fig.~\ref{fig:CI_COabun} also reveals that the observations nicely trace the actual evolution of the gas-phase CO in the model.  

In our model there is also some CO in the gas phase outside of the region affected by the burst ($r\gtrsim3000\,\mathrm{au}$, see Fig.~\ref{fig:CI_COabun}). In this region the freeze-out of CO is not efficient due to the low densities ($n_\mathrm{<H>} <10^5\,\mathrm{cm^{-3}}$) and photo-desorption. However, as the densities are low this region is optically thin in our model and does not affect the \mbox{$\mathrm{C^{18}O}\:J\!=\!2\!-\!1$} line emission from the central region. However, for some targets such a region might be more extended than in our model where the outer radius is $r=5000\,\mathrm{au}$. For such a case even the \mbox{$\mathrm{C^{18}O}\:J\!=\!2\!-\!1$} might show some self-absorption and the view towards the central region might be obscured. For such deeply embedded objects a more optically thin tracer like $\mathrm{C^{17}O}$ would be required.
\subsubsection{Recurrent bursts and initial chemical abundances}
\label{sec:initcond}
For the pre-burst initial chemical abundances we used values derived from a quiescent chemical evolution for $10^5\,\mathrm{yr}$ (Sect.~\ref{sec:outburst}). However, depending on the burst frequency, periodic bursts can alter the initial abundances. The burst-frequencies, in particular for strong and rather long lasting bursts ($\approx100\,\mathrm{yr}$) like modelled here, are not well known \citep{Audard2014}. However, models and observations indicate that accretion bursts are periodic with time spans between bursts of roughly $5000-50\,000\,\mathrm{yr}$ \citep[e.g.][]{Vorobyov2015,Scholz2013,Audard2014,Jorgensen2015b}.

For the case of quiescent periods longer than $t\approx3\times10^4\,\mathrm{yr}$ between bursts our results would not be affected at all. For such long quiescent periods the chemical abundances (at least CO) have already reached their quiescent (steady-state) levels again (see Sect.~\ref{sec:COabundance}). In case of higher burst frequencies the chemistry still will be out of equilibrium between two subsequent bursts and the pre-burst initial abundances would be different to what we have used here. 

Taking our model presented here as an example, but assuming that another burst happened about $5000\,\mathrm{yr}$ ago, the pre-burst initial CO abundances for the second burst would be similar to what is shown in the $t=3000\,\mathrm{yr}$ panel of Fig.~\ref{fig:CI_COabun}. However, if the second burst is at least as strong as the first one (here $L_\mathrm{*,B}=100\,\mathrm{L_\sun}$), CO would again sublimate out to similar radii as for the first burst and the abundance at the beginning of the post-burst phase would look the same (at least very similar) to what is shown in the model presented here. A more complicated scenario arises in the case of a weaker second burst. Fur such a case the second burst will sublimate CO only up to smaller radii than the first stronger burst and the initial post-burst abundance structure will have signatures of both bursts. As a consequence radial intensity profiles will likely show more complex shapes than what is shown here. Nevertheless, such profiles still show extended emission and can be used to identify post-burst targets. 

For the future we plan to model such a repetitive burst scenario using as input the burst properties (e.g. burst frequencies and luminosities) derived from theoretical models like presented in \citet{Vorobyov2013f,Vorobyov2015}. A detailed study of such models will allow to identify possible observational signatures of repetitive bursts with short quiescent periods.
\section{Summary and conclusions}
\label{sec:conclusions}
We have presented a new two dimensional model for the chemistry of episodic accretion in embedded objects. The model is based on the radiation thermo-chemical disk code P{\tiny RO}D{\tiny I}M{\tiny O}. We extended P{\tiny RO}D{\tiny I}M{\tiny O} with a parametric prescription for a rotating envelope structure to model a representative Class~I source consisting of a disk and envelope component. Our model features different dust size distributions for the disk (evolved dust) and envelope (ISM like dust). 

For this density structure we simulated a single burst scenario by simply increasing the quiescent luminosity by a certain factor and calculated the temperature structure and local radiation field for the quiescent and burst phase. Applying a medium sized chemical network, we followed the time evolution of the CO gas phase abundance $\mathrm{\epsilon(CO)}$ during the burst and post-burst phase. Further we presented synthetic observations (ALMA simulations) as a function of time (after the burst) for the \mbox{$\mathrm{C^{18}O}\:J\!=\!2\!-\!1$} spectral line to investigate observational signatures of the chemical evolution in the post-burst phase. Our main findings are
\begin{itemize}
  \item We used surface weighted averaged dust sizes derived from realistic dust size distributions for the disk and envelope to calculate the adsorption rate. For the disk we use grain sizes up to $1000\,\mathrm{\mu m}$, for the envelope up to $1\,\mathrm{\mu m}$. Compared to the commonly used single dust size of $0.1\,\mathrm{\mu m}$ the freeze-out timescale decreases by a factor of three in the envelope and a factor of 90 in the disk for such dust size distributions. However, as the freeze-out timescale is also a function of gas density the freeze-out timescale in the disk is typically shorter than in the envelope. As the density decreases as a function of distance from the protostar, the freeze-out of e.g. CO happens from inside (high densities) out (low densities).
  \item Including a disk component has a significant impact on the temperature structure of the envelope. Due to the disk shadow the temperatures close to the midplane of the envelope are cooler compared to structures without a disk. In contrast to a model without a disk, the average CO abundance within the radial CO ice line of the envelope is therefore depleted (by a factor of three in our model). Such effects can not be properly modelled by 1D models. However, on large scales the freeze-out chemistry is mainly driven by the density gradient in the envelope, which is not affected by the disk. Therefore the CO gas phase evolution in the post-burst phase is similar to structures without a disk component.
  \item The synthetic \mbox{$\mathrm{C^{18}O}\:J\!=\!2\!-\!1$} ALMA observations show that the inside-out freeze-out produces distinct observational signatures. The most striking feature is a clearly visible gap in the intensity images which is caused by the differences in the freeze-out timescales between the zone close to the quiescent CO ice line and the zone close to the burst CO ice line. Such a gap is likely only visible if the target is seen nearly face-on, when one can peek down the outflow cavity. For inclined targets such a gap is not visible. However, the inside-out freeze-out still has an impact on the intensity maps (X-shaped emission pattern) and  radial intensity profiles. The peak intensity of the radial profiles drops on shorter timescales than the extended emission.
  \item Based on our models we propose a new method to identify post-burst targets via spatially resolved $\mathrm{C^{18}O}$ observations of embedded objects. The $\mathrm{C^{18}O}$ emission in the post-burst phase consists of two components, one corresponds to the centrally peaked emission, which also exists during the quiescent phase, the second component corresponds to the extended emission which only exists for a limited time (up to $10\,000\,\mathrm{yr}$) after the burst. The post-burst emission pattern is much better fitted by a two Gaussian fit where the quiescent emission pattern is better matched with a single Gaussian. A successful two Gaussian fit is therefore already an indication for a recent burst. This method is model independent and in particular does not depend on the CO binding energy.
  \item Our model results confirm that measuring the extent of $\mathrm{CO}$ emission in  embedded sources \citep{Vorobyov2013f,Jorgensen2015b} is an efficient method to identify post-burst objects up to $\approx10\,000\,\mathrm{yr}$ after the a burst. However, to derive reliable statistical properties such as burst frequencies from an observational sample, the possible different structures and inclinations of the individual targets should be taken into account.
\end{itemize}
\begin{acknowledgements}
We would like to thank Ruud Visser for providing a detailed and clear referee's report that helped to improve the paper. We also want to thank Zhaohuan Zhu for providing the burst spectrum of FU Orionis. RCH and EV acknowledge funding by the Austrian Science Fund (FWF): project number  I2549-N27. RCH and MG acknowledge funding by the Austrian Science Fund (FWF): project number P24790. MA and AP acknowledge funding by the Swiss National Science Foundation (SNSF): project number: 200021L\_163172. The research leading to these results has received funding from the European Union Seventh Framework Programme \mbox{FP7-2011} under grant agreement no 284405. This publication was partly supported from the FFG ASAP 12 project \textit{JetPro*} (FFG-854025). This publication was supported by the Austrian Science Fund (FWF). The computational results presented have been achieved using the Vienna Scientific Cluster (VSC). This research has made use of NASA's Astrophysics Data System. All figures were made with the free Python module Matplotlib \citep{Hunter2007}. This research made use of Astropy, a community-developed core Python package for Astronomy \citep{AstropyCollaboration2013}.
\end{acknowledgements}
\bibliographystyle{aa}
\bibliography{Rab_EpisodicAccretionChemistry}
\appendix
\section{Comparison to \citet{Visser2015c}}
\label{sec:visser}
\begin{table}
\caption{Model parameters for the comparison with \citet{Visser2015c}.}
\label{tab:EVmodel}
\centering
\begin{tabular}{l|c|c}
\hline\hline
Quantity & Symbol & Value \\
\hline
stellar mass                          & $M_\mathrm{*}$                    & $0.5~\mathrm{M_{\sun}}$\\
stellar effective temp.               & $T_{\mathrm{*}}$                  & 5000~K\\
stellar luminosity                    & $L_{\mathrm{*}}$                  & $1.6\,\mathrm{L_{\sun}}$\\
\hline
strength of interst. FUV              & $\chi^\mathrm{ISM}$               & 1\tablefootmark{a}\\
cosmic ray $\mathrm{H_2}$ ion. rate   & $\mathrm{\zeta_{CR}}$             & $\mathrm{5\times10^{-17}\,s^{-1}}$\\
\hline
centrifugal radius                    & $R_{\mathrm{c}}$               & 0~au \tablefootmark{b}\\
mass infall rate                   & $\dot{M}_\mathrm{if}$                         & $2\times10^{-5}\,\mathsf{M_\sun yr^{-1}}$\\
inner radius                          & $R_{\mathrm{E,in}}$               & 6.2~au\\
outer radius                          & $R_{\mathrm{E,out}}$              & 6200~au\\
cavity opening angle                  & $\beta_\mathrm{cav}$              & $20^{\circ}$ \\
dust/gas mass ratio                   & $d/g$                             & 0.01\\
\hline
min. dust particle radius             & $a_\mathrm{min}$                  & $\mathrm{0.005~\mu m}$  \\
max. dust particle radius             & $a_\mathrm{max}$                  & $\mathrm{0.25~\mu m}$  \\
dust size dist. power index           & $a_\mathrm{pow}$                  & 3.5 \tablefootmark{c}\\
\hline
\end{tabular}
\tablefoot{
\tablefoottext{a}{$\chi^\mathrm{ISM}$ is given in units of the Draine field \citep{Draine1996b,Woitke2009a}.}
\tablefoottext{b}{Using $R_\mathrm{c}=0$ results in a spherical symmetric density distribution.}
\tablefoottext{c}{For the dust composition see Table~\ref{table:model}}
}
\end{table}
\begin{table}
\caption{Initial abundances for the time-dependent chemistry. }
\center
\label{tab:tcheminit}
\begin{tabular}{lr}
\hline \hline
Species & $\epsilon(X)$  \\
\hline
H      & 5.0(-5) \\
H$_2$  & 0.5       \\
He     & 0.09  \\
CO     & 5.17(-5)  \\
CO\#    & 4.89(-5)  \\
CO$_2$\#   & 3.91(-5)  \\
N      & 4.12(-5)  \\
N$_2$  & 1.57(-5)  \\
NH$_3$\#& 2.25(-6)  \\
H$_2$O\#& 1.41(-4)  \\
S      & 8.00(-8)  \\
Si     & 8.00(-9)  \\
Na     & 2.00(-9)  \\
Mg     & 7.00(-9)  \\
Fe     & 3.00(-9)  \\
\hline
\end{tabular}
\tablefoot{These are the same abundances as in \citet{Visser2015c}}
\end{table}
\begin{table}
\caption{Adsorption Energies of key molecules.}
\center
\label{tab:adsenergies}
\begin{tabular}{lrl}
\hline \hline
Species & $\mathrm{E_B\,[K]}$ \\
\hline
O      & 1420\tablefootmark{1,2} \\ 
CO     & 1307\tablefootmark{3}~~ \\
CO$_2$ & 2300\tablefootmark{3}~~ \\
H$_2$O & 5773\tablefootmark{4}~~ \\
N$_2$  & 1200\tablefootmark{5}~~ \\
NH$_3$ & 2790\tablefootmark{6}~~ \\
\hline
\end{tabular}
\tablefoot{\tablefoottext{1}{\citep{Minissale2015}}, \tablefoottext{2}{\citep{Minissale2016a}}, \tablefoottext{3}{\citep{Noble2012b}}, \tablefoottext{4}{\citep{Fraser2001}}, \tablefoottext{5}{\citep{Visser2015c,Fayolle2016}},\tablefoottext{6}{\citep{Brown2007p}}. The adsorption energies for all other species are taken from the UMIST 2012 release \citep{McElroy2013b} \url{http://udfa.ajmarkwick.net/downloads/RATE12_binding_energies.dist.txt}}
\end{table}
In this Section we present a comparison of our model and the 1D model of \citet{Visser2015c}. The main goal of this comparison is to verify our chemical model, as we use in contrast to \citet{Visser2015c} a rather small chemical network. For the comparison we chose their model for the low-mass protostar \object{IRAS 15398}. For that model \citet{Visser2015c} presented detailed density, temperature and molecular abundance profiles.

\citet{Visser2015c} uses the 1D spherical density and temperature profiles from the \emph{DUSTY} \citep{Ivezic1997d} radiative transfer models of \citet{Kristensen2012l,Jorgensen2002q}. The outburst is modelled by increasing the quiescent stellar luminosity of $L_{\mathrm{*}}=1.6\,\mathrm{L_{\sun}}$ by a factor of 100. Their chemical network is based on the 2012 release of the UMIST Database for Astrochemistry \citep{McElroy2013b}. In addition they include adsorption and desorption for all neutral molecules, formation of H$_2$ on dust grains and hydrogenation of C, N and O on dust grain surfaces.   

For the density structure we use the same 2D description for the envelope structure as discussed in Sect.~\ref{sec:struc}. To achieve a similar density distribution as used by \citet{Visser2015c}, a power-law with $\rho(r)\propto r^{-1.4}$, we set the centrifugal radius $R_\mathrm{c}$ in Eq.~\ref{eqn:envstruc3} to zero and do not include a disk component. This provides a spherical symmetric density distribution following a radial power-law with $\rho(r)\propto r^{-1.5}$ (i.e. slightly steeper than the profile used by \citealt{Visser2015c}). The model still includes an outflow cavity with an opening angle of 20\degr.

\begin{figure}
\centering
\includegraphics{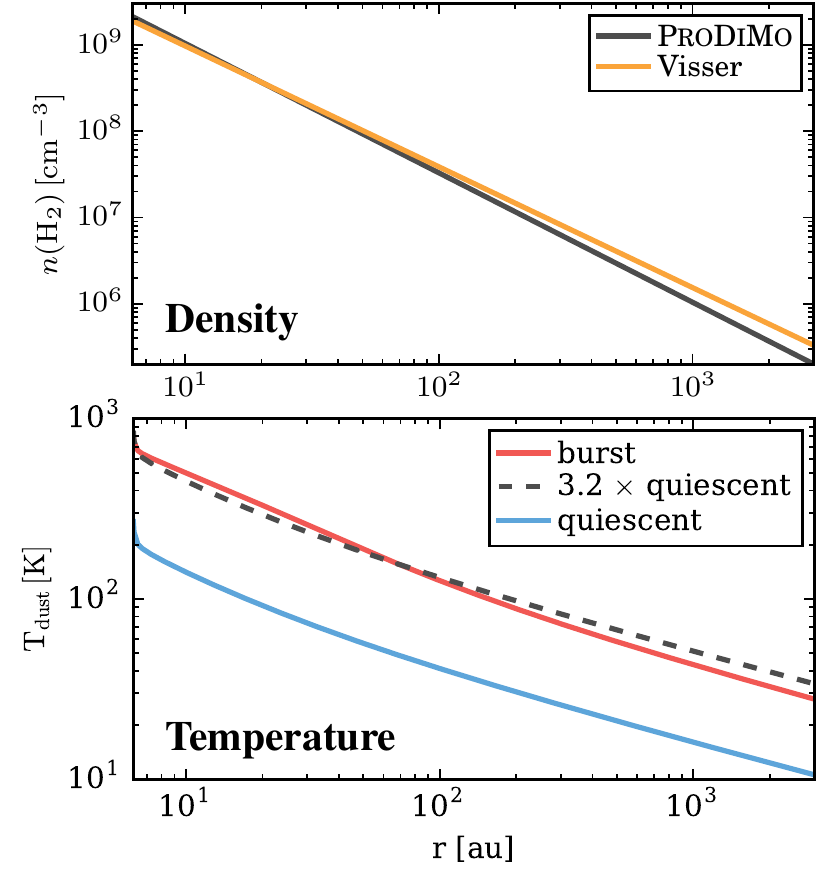}
\caption{Radial midplane density and temperature profiles of the model used for the comparison to the 1D model of \citet{Visser2015c}. In the molecular hydrogen density plot (top panel) also the density profile of \citet{Kristensen2012l,Visser2015c} is shown. In the temperature plot (bottom panel) the red solid line represents the burst phase and the blue solid line the quiescent period. The dashed grey line shows the quiescent temperature profile multiplied by a factor of 3.2.}
\label{fig:visserStruc}
\end{figure}
\begin{figure*}
\sidecaption
\includegraphics{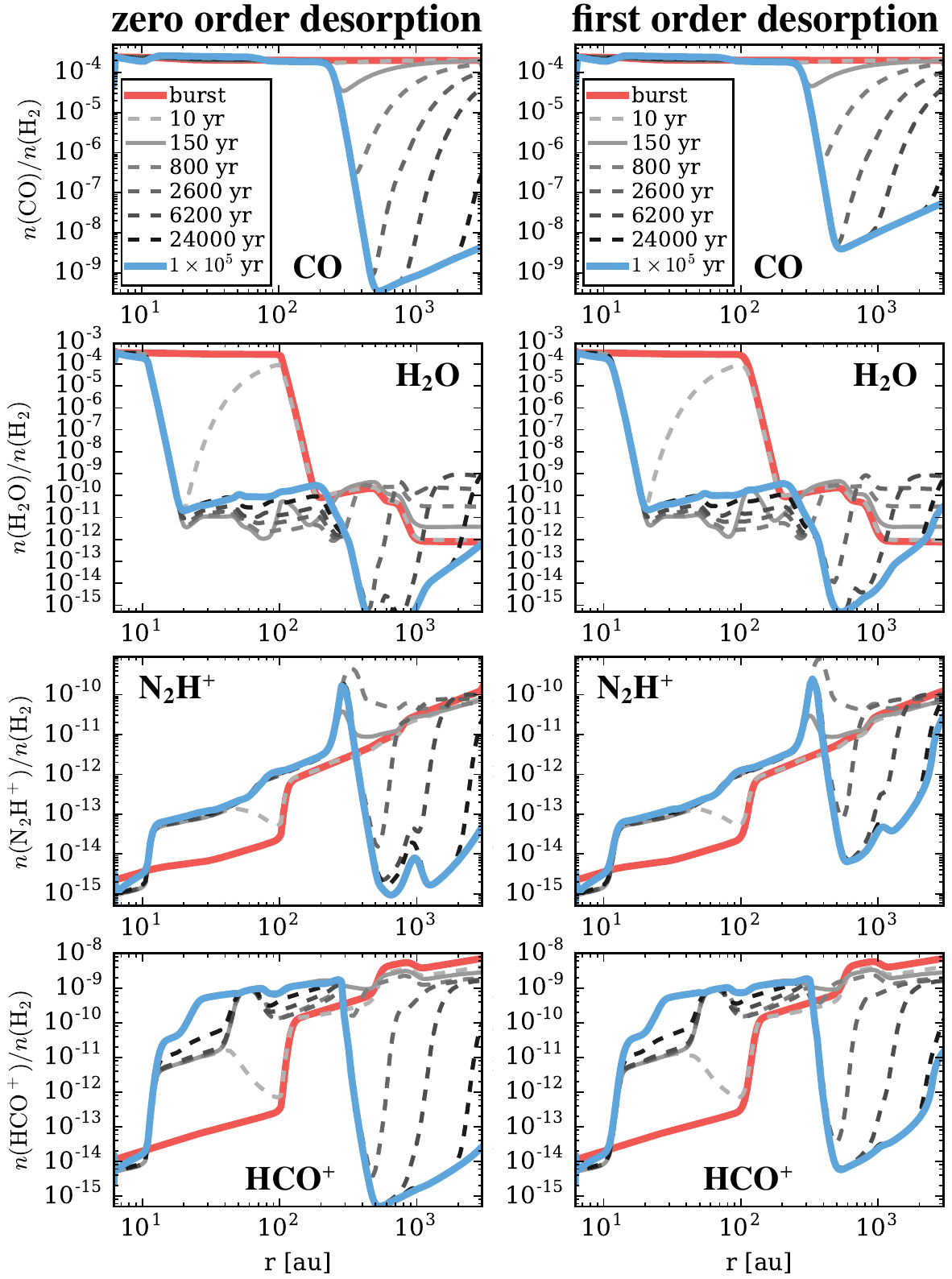}
\caption{Radial midplane abundance profiles for the model used for the comparison to \citet{Visser2015c}. In contrast to the rest of the paper the abundances are  given relative to molecular hydrogen like in \citet{Visser2015c}. Shown are, from top to bottom, CO, H$_2$O, N$_2$H$^+$ and HCO$^+$ during the burst (red solid line) and at different times in the post-burst phase (grey and blue solid lines). The left column shows a model with zeroth order desorption (P{\tiny RO}D{\tiny I}M{\tiny O} standard) the right column a model using first order desorption. This Figure can be compared to Fig.~1 of \citet{Visser2015c}}. 
\label{fig:visser}
\end{figure*}
For the stellar properties and the burst we used the same parameters as \citet{Visser2015c}. The effective temperature $T_\mathrm{*}$ of the star is not given in \citet{Visser2015c}. However, as their dust continuum model is based on \citet{Kristensen2012l,Jorgensen2002q} we assumed $T_\mathrm{*}=5000\,\mathrm{K}$ as given in \citet{Jorgensen2002q}. We have not not use the same dust opacities as \citet{Kristensen2012l}, who use the opacities from \mbox{\citet{Ossenkopf1994a}}. We applied the same dust composition as in the main paper (see Sec.~\ref{sec:dustprop}) and only adapt the dust size distribution. All relevant physical parameters for the comparison model are summarized in Table~\ref{tab:EVmodel}.

As we used a 2D model we only compare the midplane (i.e. the plane perpendicular to the outflow axis) quantities of our model to the 1D model of \citet{Visser2015c}. In the midplane the differences caused by the different structures used (e.g. outflow cavity) should be minimal. In Fig.~\ref{fig:visserStruc} we show the radial density and temperature profiles in the midplane. In the plot for the density also the power-law density profile used by \citet{Visser2015c} is shown for comparison.
As we used a two-dimensional structure, a slightly different density profile, and different dust opacities, it is not surprising that there are some deviations in the dust temperatures compared the model of \citet{Visser2015c}. In the quiescent state our model gives a dust temperature at the inner radius of $T_\mathrm{d}(R_\mathrm{E,in})=269\,\mathrm{K}$ and a radius where the dust temperature reaches $10\,\mathrm{K}$ of $r(T_\mathrm{d}=10\,\mathrm{K})\approx3400\,\mathrm{au}$, compared to $T_\mathrm{d}(R_\mathrm{E,in})=250\,\mathrm{K}$ and $r(T_\mathrm{d}=10\,\mathrm{K})\approx2700\,\mathrm{au}$ of \citet{Visser2015c}.

Consistent with \citet{Visser2015c} the temperature increases throughout the envelope during the outburst. According to \citet{Johnstone2013} this increase follows roughly  $T_\mathrm{d}\propto L_\mathrm{*}^{0.25}$. For an increase of the luminosity by a factor of $100$ this corresponds to an increase of $T_\mathrm{d}$ by a factor of $\approx3.2$. This is also the case in our model as as seen in Fig.~\ref{fig:visserStruc}.

To compare the results of our chemical model with \citet{Visser2015c} we adapted their initial chemical abundances (Table~\ref{tab:tcheminit}) and their adsorption energies for key molecules (Table~\ref{tab:adsenergies}). All other chemical model parameters (e.g. sticking coefficient) are left unchanged (see Sect.~\ref{sec:chemmodel} for details). In Fig.~\ref{fig:visser} we show radial midplane abundance profiles for the molecules CO, H$_2$O, N$_2$H$^+$ and HCO$^+$ during the burst and at several times  after the burst (post-burst phase). The Figure shows the results for a model using zeroth order desorption and a model where we used first order desorption. Fig.~\ref{fig:visser} can be directly compared to Fig.~1 in \citet{Visser2015c}.  

In general our chemical model results are in good agreement with \citet{Visser2015c}. Our model nicely reproduces the main aspects of episodic accretion chemistry, namely the delayed freeze-out of neutral gas phase molecules and the outward shift of ice lines (here shown for CO and H$_2$O). Also the profiles for ions (N$_2$H$^+$ and HCO$^+$) are in good agreement with \citet{Visser2015c}. The chemistry of the ions is  strongly sensitive to the gas phase abundances of the neutrals as the ions are efficiently destroyed by reactions with H$_2$O and/or CO. As a consequence N$_2$H$^+$ and HCO$^+$ trace the ice lines of H$_2$O and CO (see \citealt{Visser2012m,Visser2015c} for details).    

However, we also find differences in particular for the quiescent abundance profiles ($t=10^5\,\mathrm{yr}$ in Fig.~\ref{fig:visser}). The minimum abundances in the quiescent state for all shown molecules are about an order of magnitude lower (even more for water) than in the model of \citet{Visser2015c}. Further, at radii $\gtrsim1000\,\mathrm{au}$ the ion abundances are more than a factor of $100$ below the values found by \citet{Visser2015c}. To some extent this deviation can be explained by the differences in the structure and radiative transfer model. However, we find that the abundances profiles are not very sensitive to changes in, for example, the density profile, and the quiescent abundances change at most by a factor of a few. 

Most likely the differences are caused by the details of the model for the adsorption and desorption processes. Many different parameters like the sticking coefficient, desorption yields and the average dust sizes (see Sect.~\ref{sec:chemmodel}) are relevant. However, we find that actually the treatment of the thermal desorption process is most relevant. Comparing the results for the zeroth order desorption and first order desorption in Fig.~\ref{fig:visser}, clearly shows an increase of the minimum abundances by an order of magnitude if first order desorption is used (see Sect.~\ref{sec:chemmodel}). 

The above discussed results show that for a spherical symmetric structure, our model is in good agreement with the model of \citet{Visser2015c} for both the dust radiative transfer and the chemistry. In particular the two models agree very well concerning the main aspect of episodic accretion chemistry which is the delayed freeze-out of neutral species in the post-burst phase.
\section{ALMA/CASA Simulations}
\label{sec:alma_method}
To produce as realistic as possible synthetic observations we use the Common Astronomy Software Applications (CASA) package \citep{McMullin2007}. We use CASA to either convolve the spectral line cubes with an artificial beam or to run full ALMA  simulations. For the azimuthally averaged radial intensity profiles we use the task \texttt{casairing} provided by the Nordic ALMA regional centre. 

\begin{figure}
\centering
\resizebox{\hsize}{!}{\includegraphics{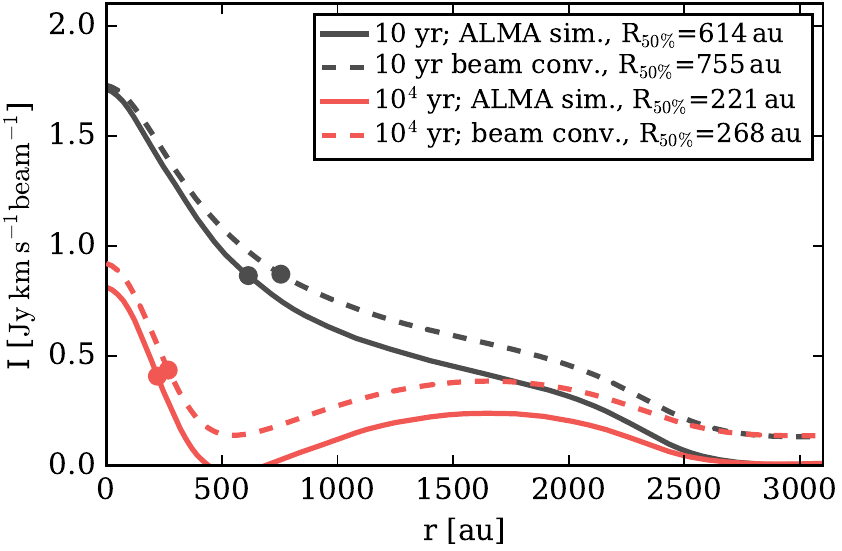}}
\caption{Comparison of full ALMA simulations to beam convolved models (same beam size). Shown are azimuthally averaged \mbox{$\mathrm{C^{18}O}\:J\!=\!2\!-\!1$} radial intensity profiles at 10 and 10$^4\,\mathrm{yr}$ after the burst. The solid lines are for full ALMA simulations, the dashed lines are for the beam convolved models.}
\label{fig:radp_csim}
\end{figure}

In the following we describe in detail the main steps for the ALMA simulations used to produce \mbox{$\mathrm{C^{18}O}\:J\!=\!2\!-\!1$} (ALMA Band 6) line images and radial profiles.
\begin{enumerate}
  \item We performed spectral line transfer with P{\tiny RO}D{\tiny I}M{\tiny O} to produce line cubes with 101 velocity channels (spectral resolution of $0.1\,\mathrm{km\,s^{-1}}$);
  \item We used the most compact 12m Array configuration (full operations) in combination with the ACA (Atacama Compact Array) to cover all spatial scales. The observation time for the full array is $2\,\mathrm{h}$ and for the ACA $10\,\mathrm{h}$ (a factor of five longer than for the 12m Array, as recommend in the ALMA proposal guide for Cycle~4). The maximum recoverable scale for the full Array and ACA are $12.6\arcsec$ and $29.0\arcsec$ ($2520\,\mathrm{au}$ and $5800\,\mathrm{au}$ at a distance of $200\,\mathrm{au}$), respectively. The observations are simulated with the task \texttt{simobserve} (including noise);
  \item With the tasks \texttt{concat} and \texttt{split} we combined the observations and rebined the line cube by a factor of five, resulting in 20 channels with a spectral resolution of $\approx0.5\,\mathrm{km\,s^{-1}}$;   
  \item We performed continuum subtraction in the visibility plane using the task \texttt{uvcontsub};
  \item We reconstructed the images with \texttt{simanalyze} applying a threshold of $1\,\mathrm{mJy}$ and \emph{briggs} weighting. The resulting synthetic beam size is approximately $1.81\arcsec\times 1.65\arcsec$ ($362\,\mathrm{au}\times 330\,\mathrm{au}$ at a distance of $200\,\mathrm{au}$) and the root mean square noise of the images is $rms \lesssim 0.04\mathrm{\,Jy\:beam^{-1}}$;  
  \item We generated moment zero maps (intensity maps) with the task \texttt{immoments};
  \item We generated azimuthally averaged radial intensity profiles with \texttt{casairing} (Nordic ALMA regional centre).
 \end{enumerate}
We also performed simulations where the line cubes of P{\tiny RO}D{\tiny I}M{\tiny O} are simply convolved with an elliptical beam of the same size as is used for the ALMA simulations. A comparison of the simple beam convolved simulations to the full ALMA simulations shows that we loose about $30\%$ of the total flux in the ALMA simulations. This is also seen in Fig.~\ref{fig:radp_csim} where we show a comparison of radial intensity profiles for the full ALMA simulations and the beam convolved simulations. However, Fig.~\ref{fig:radp_csim} also shows that the full ALMA simulations recover the main spatial features of the profile very well but miss some flux in particular at larger scales. The cause of this might be an insufficient coverage of spatial scales and/or a inperfect image reconstruction (cleaning).
\section{ALMA simulations for inclined models}
\begin{figure*}
\centering
\resizebox{0.973\hsize}{!}{\includegraphics{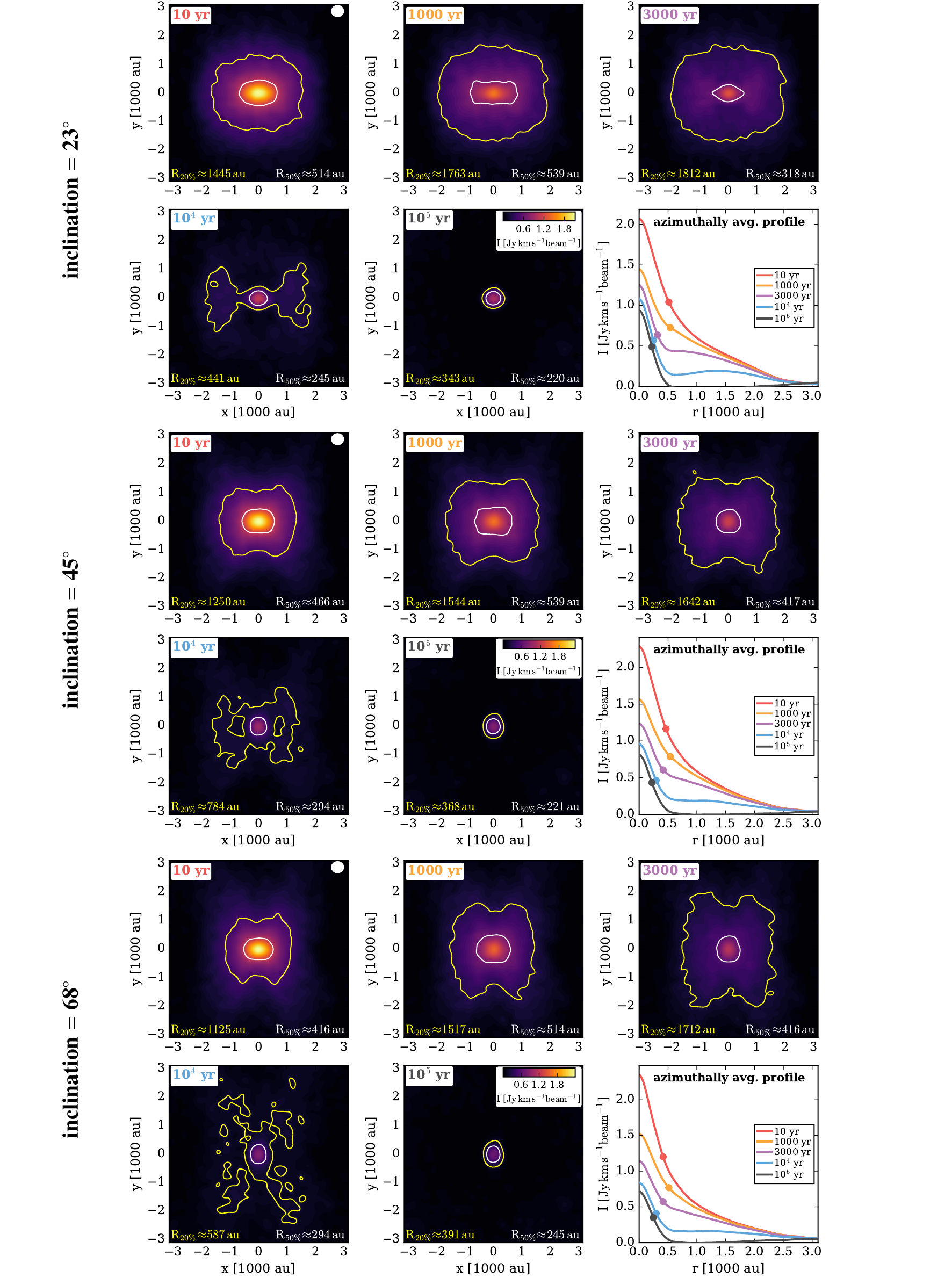}}
\caption{Same as Figs.~\ref{fig:almasim0} and \ref{fig:almasim90}  but for inclinations of 23\degr, 45\degr and 68\degr (from top to bottom).}
\label{fig:almasiminc}
\end{figure*}

In Fig.~\ref{fig:almasiminc} we show the same \mbox{$\mathrm{C^{18}O}\:J\!=\!2\!-\!1$} ALMA simulation as shown in Figs.~\ref{fig:almasim0} and \ref{fig:almasim90} but for inclinations of 23\degr, 45\degr and 68\degr. Fig.~\ref{fig:almasiminc} shows that the gap in the \mbox{$\mathrm{C^{18}O}\:J\!=\!2\!-\!1$} emission is only visible for weakly inclined targets and that the X-shape of the emission is most pronounced for strongly inclined targets (see Sect.~\ref{sec:almasim} for details).
\section{Radial intensity profiles for the structure models}
\begin{figure*}
\centering
  \resizebox{0.95\hsize}{!}{\includegraphics{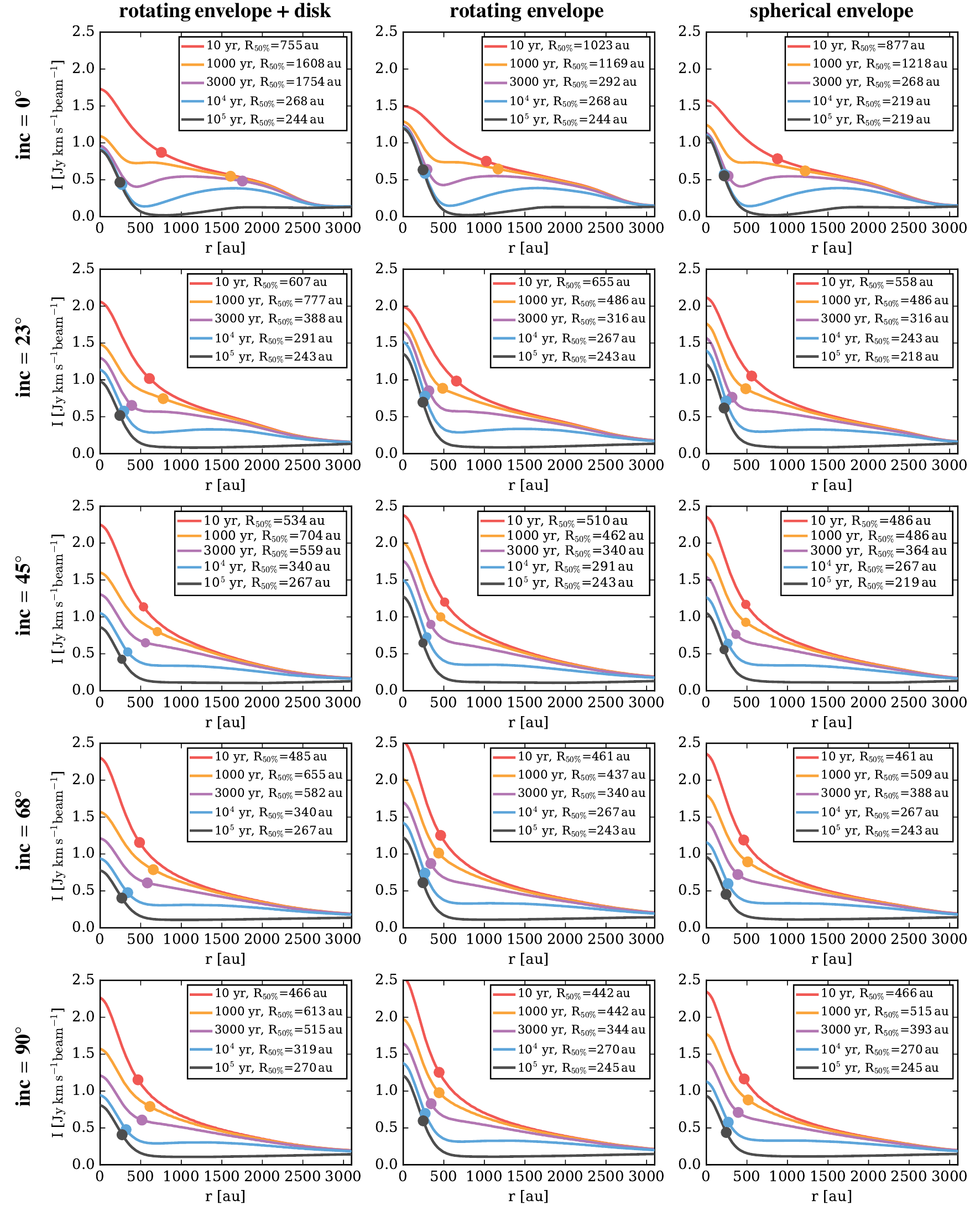}}  
  \caption{Time evolution of the azimuthally averaged \mbox{$\mathrm{C^{18}O}\:J\!=\!2\!-\!1$} radial intensity profiles for the three different structure models (columns) viewed at five different inclinations (rows). See also Fig.~\ref{fig:struc_rp} for details.}
  \label{fig:rp_struciall}
\end{figure*}
In Fig.~\ref{fig:rp_struciall} we show radial \mbox{$\mathrm{C^{18}O}\:J\!=\!2\!-\!1$} intensity profiles for our structure models (see Fig.~\ref{fig:struc_rp}) but for all five inclinations considered for the synthetic observations. 
\section{Further fitting examples}
\label{sec:fittingexamples}
\begin{figure*}
\centering
  \resizebox{0.80\hsize}{!}{\includegraphics{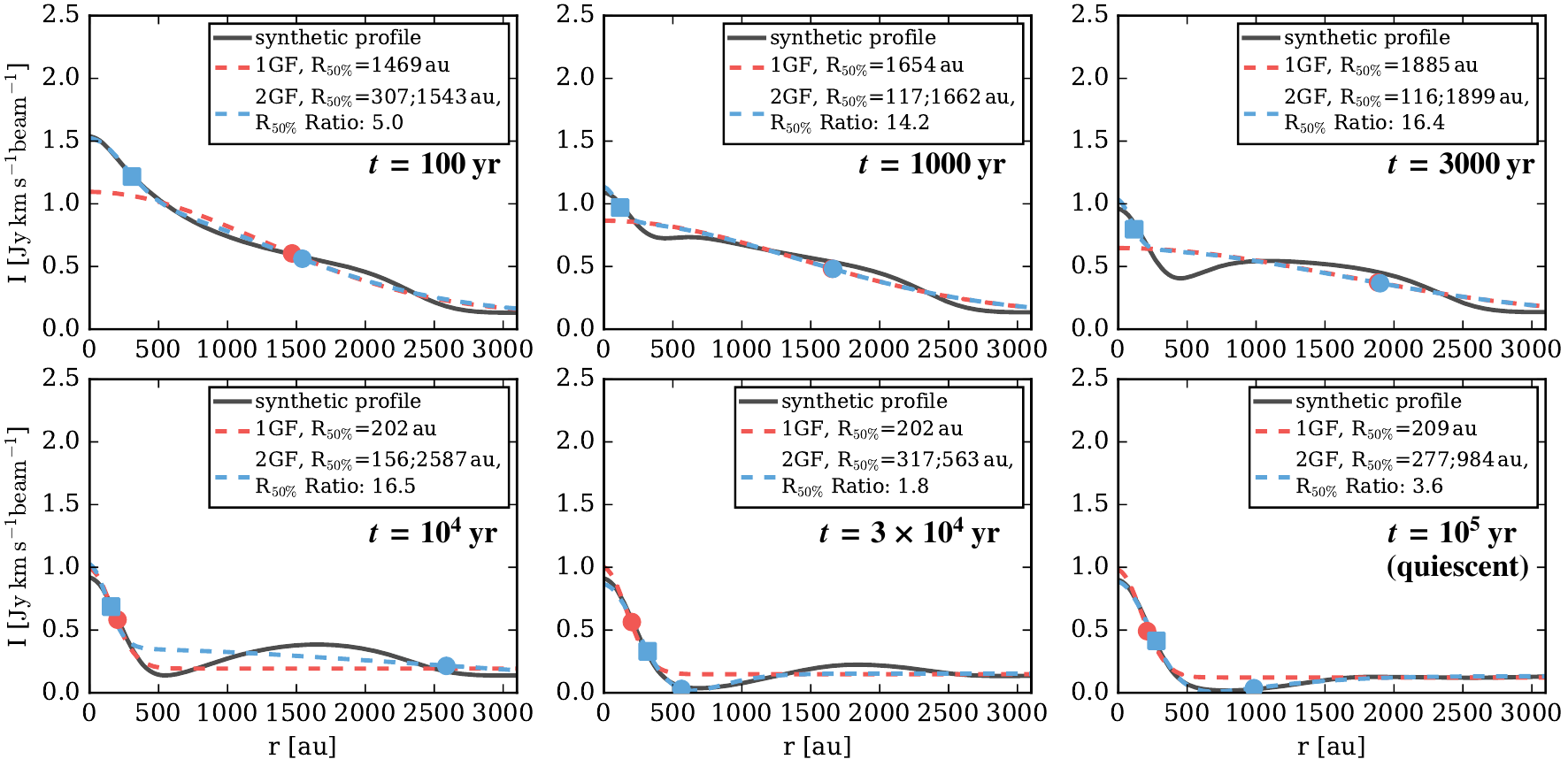}}  
  \caption{Same as Fig.~\ref{fig:almafit} but for a model with an inclination of 0\degr (face-on).}
  \label{fig:almafiti0}
\end{figure*}

\begin{figure*}
\centering
  \resizebox{0.80\hsize}{!}{\includegraphics{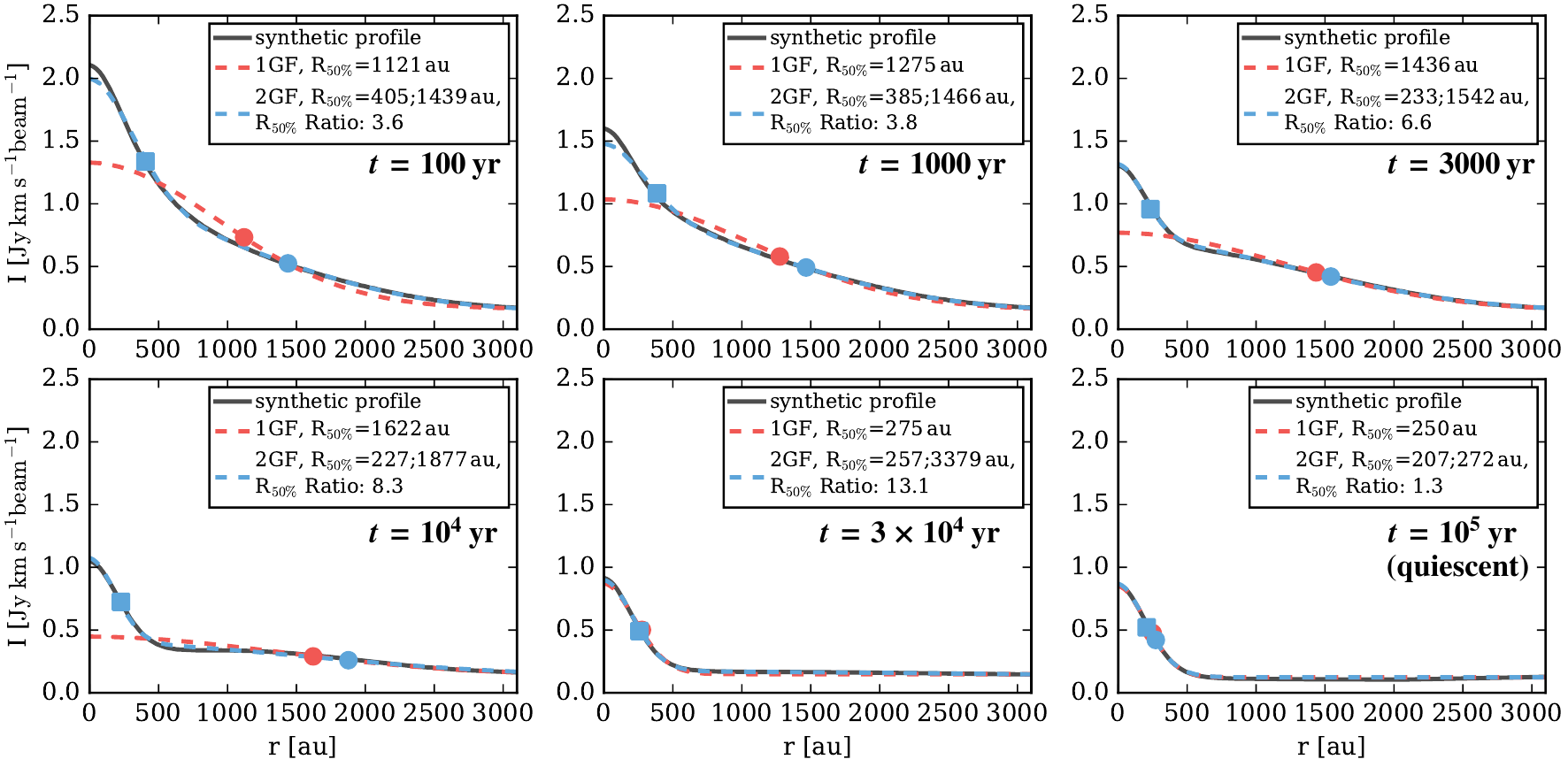}}  
  \caption{Same as Fig.~\ref{fig:almafit} but for a model with a CO binding energy of 960K (CO sublimation temperature of $\approx 20\,\mathrm{K}$). We note the two Gaussian fit (2GF) for $t=3\times10^4\,\mathrm{yr}$ did not converge for this particular model.}
  \label{fig:almafitEB20}
\end{figure*}

\begin{figure*}
\centering
  \resizebox{0.80\hsize}{!}{\includegraphics{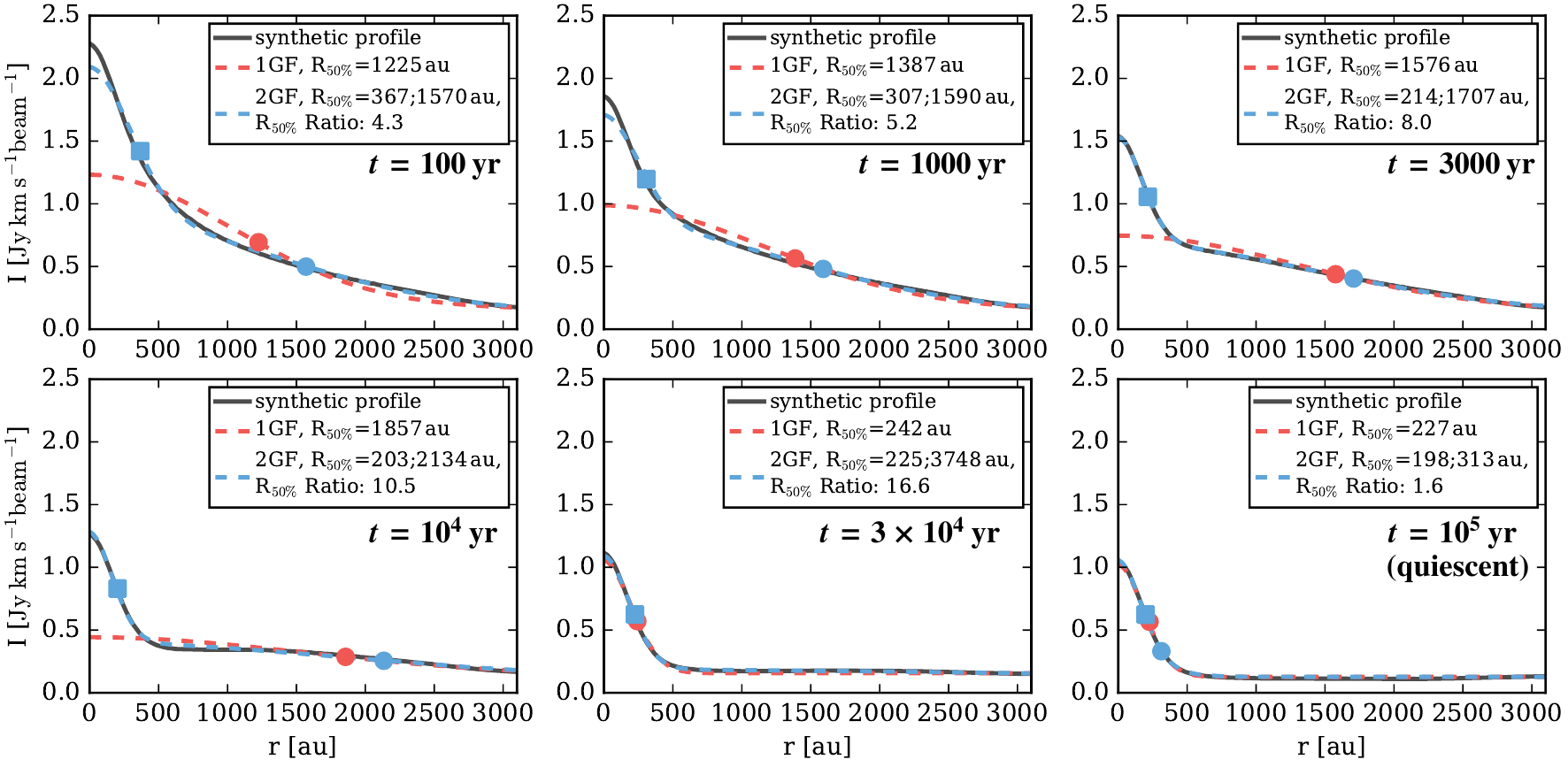}}  
  \caption{Same as Fig.~\ref{fig:almafit} but for a model with a spherical density distribution (see Sect.~\ref{sec:impactstruc}).}
  \label{fig:almafitSPH}
\end{figure*}

\begin{figure*}
\centering
  \resizebox{0.80\hsize}{!}{\includegraphics{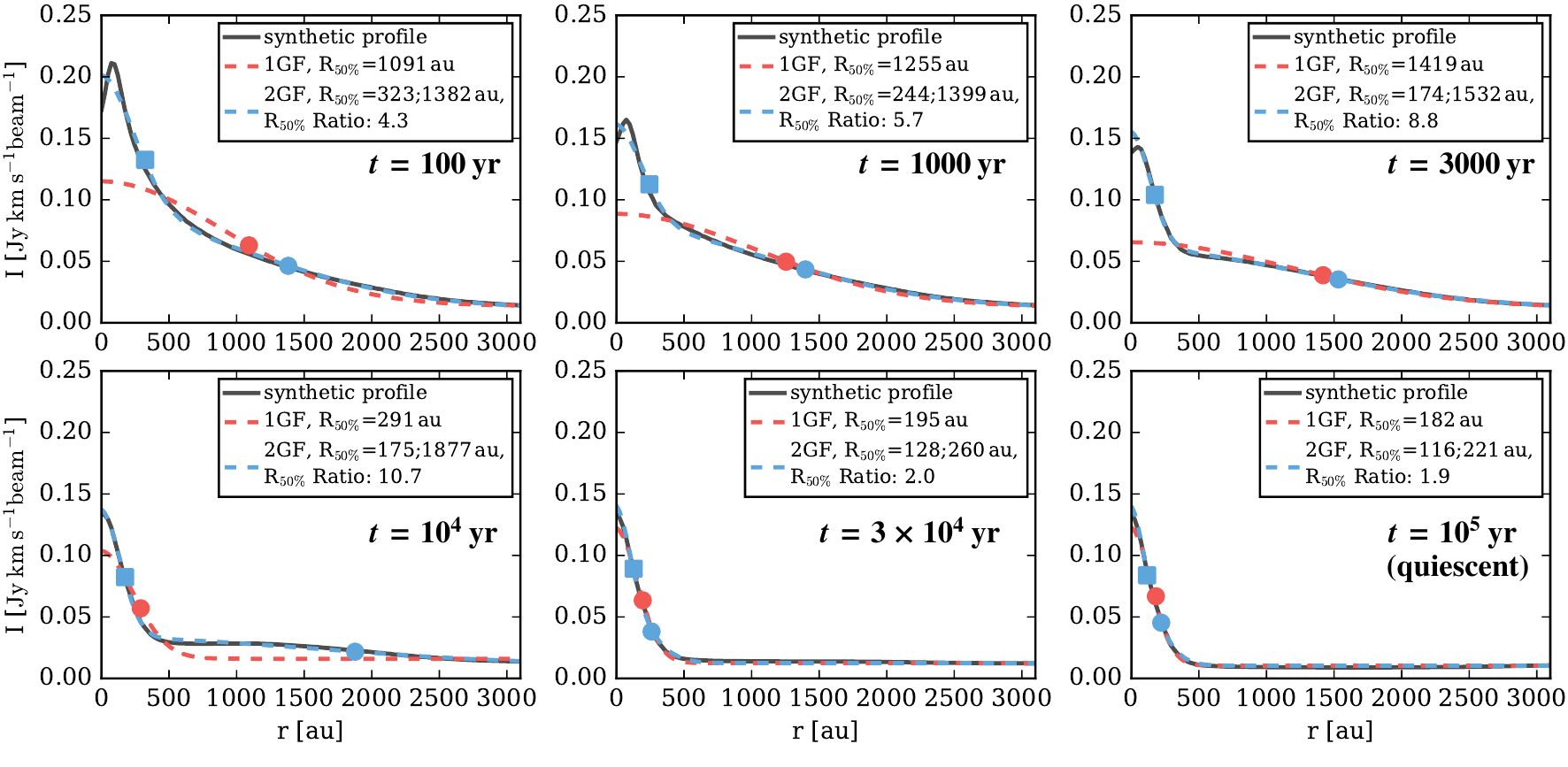}}  
  \caption{Same as Fig.~\ref{fig:almafit} but for a model with a synthetic beam of $0.5\arcsec\times0.5\arcsec$.}
  \label{fig:almafitHR}
\end{figure*}

\begin{figure*}
\centering
  \resizebox{0.8\hsize}{!}{\includegraphics{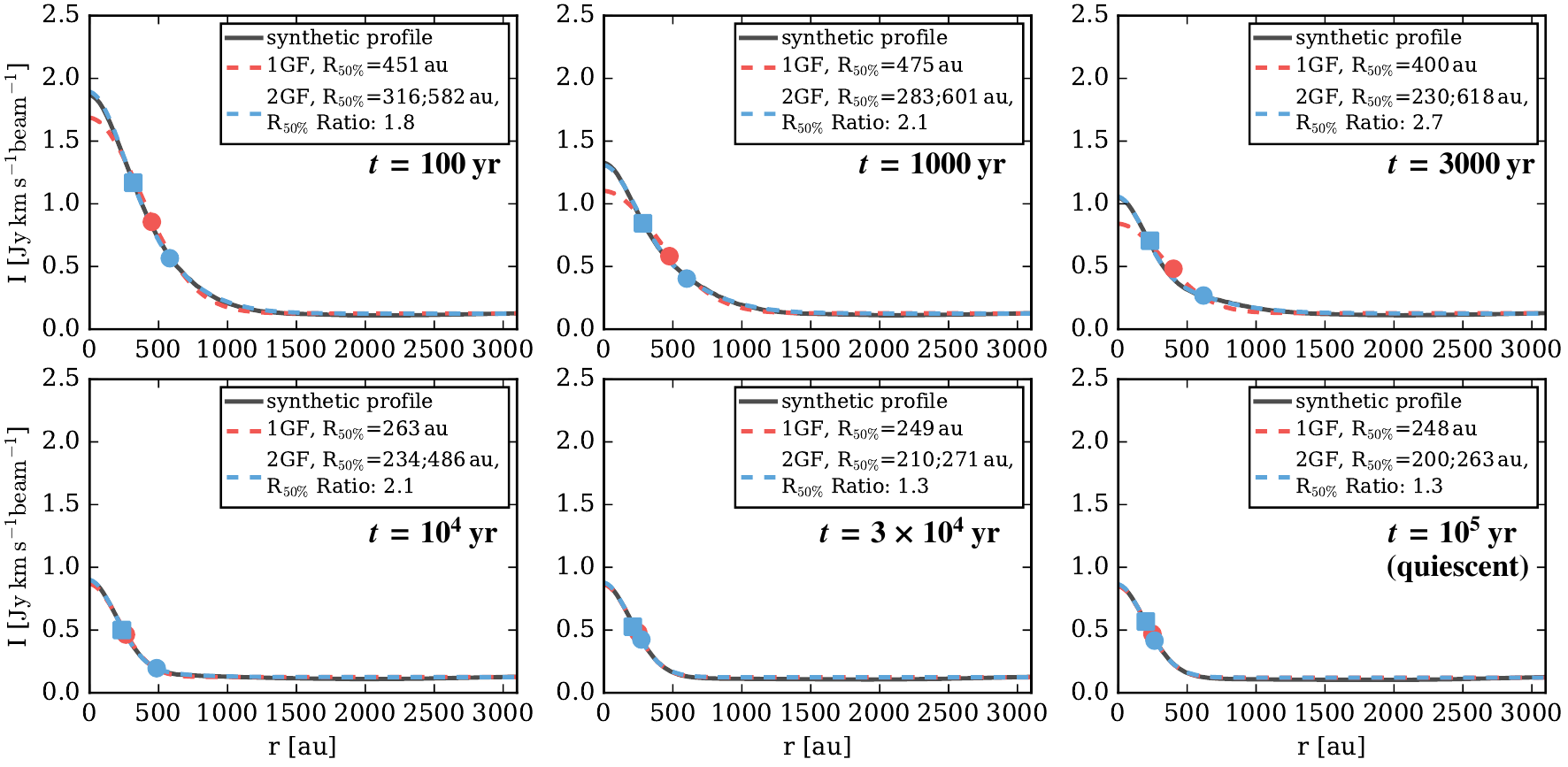}}  
  \caption{Same as Fig.~\ref{fig:almafit} but for a model with a weak burst of $10\times L_\mathrm{\mathrm{*}}$.}
  \label{fig:almafit10L}
\end{figure*}

\begin{figure*}
\centering
  \resizebox{0.8\hsize}{!}{\includegraphics{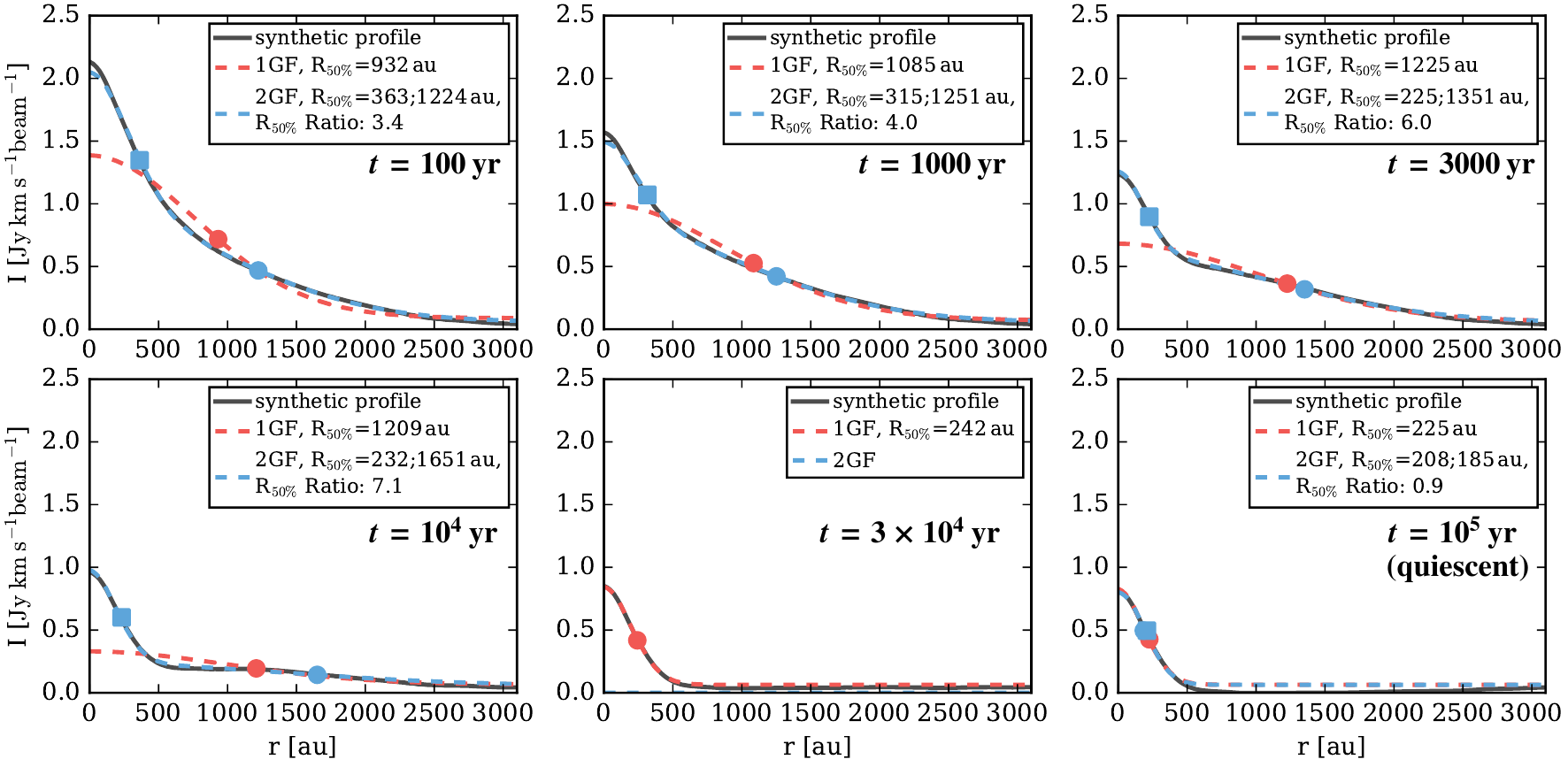}}  
  \caption{Same as Fig.~\ref{fig:almafit} but for the full ALMA simulations with an inclination of 45\degr. For the corresponding images see Fig.~\ref{fig:almasiminc}.}
  \label{fig:almaalma}
\end{figure*}

To test the robustness of our fitting procedure discussed in Sect.~\ref{sec:fitdiscussion} we applied the procedure to models with an inclination of 0\degr~(Fig.~\ref{fig:almafiti0}), a CO binding energy of $960\,\mathrm{K}$ (Fig.~\ref{fig:almafitEB20}), a spherical structure without a disk (Fig.~\ref{fig:almafitSPH}), a higher spatial resolution (Fig.~\ref{fig:almafitHR}), a weaker burst (Fig.~\ref{fig:almafit10L}) and a model using the full ALMA simulations (Fig.~\ref{fig:almaalma}). These figures show that the method for identifying post-burst target outlined in Sect.~\ref{sec:fitdiscussion} is not strongly sensitive to model properties such as inclination, structure, spatial resolution of the observations, and the actual CO binding energy.
\end{document}